\documentclass[preprint,12pt]{elsarticle}

\usepackage{amssymb}
\usepackage{amsmath}
\usepackage{graphicx}
\usepackage{dcolumn}	% Align table columns on decimal point
\usepackage{bm}			% bold math
\usepackage{color}
\usepackage{url}
\usepackage{subfigure}
\journal{Physica A}

\begin{document}

\begin{frontmatter}
%-----------------------------------------------------------------
\title{On possible origins of trends \\
       in financial market price changes}
%-----------------------------------------------------------------
%	Address
%----------------------
\author[a1]{Ryo Murakami}
%----------------------
\author[a1]{Tomomichi Nakamura}
\ead{tomo@sim.u-hyogo.ac.jp}
%----------------------
\author[a1]{Shin Kimura}
%----------------------
\author[a2]{Masashi Manabe}
%----------------------
\author[a3]{Toshihiro Tanizawa}
\ead{tanizawa@ee.kochi-ct.ac.jp}
%----------------------
\address[a1]{Graduate School of Simulation Studies, University of Hyogo,\\
             7-1-28 Minatojima-minamimachi, Chuo-ku, Kobe, Hyogo 650-0047, Japan}
%-----------------------------------------------------------------
\address[a2]{Faculty of Business Innovation, Kaetsu University, \\
             2-8-4 Minami-cho, Hanakoganei, Kodaira-shi, Tokyo 187-8578, Japan}
%	https://www.kaetsu.ac.jp/
%-----------------------------------------------------------------
\address[a3]{Kochi National College of Technology, \\
             200-1 Monobe-Otsu, Nankoku, Kochi 783-8508, Japan}
%-----------------------------------------------------------------
\begin{abstract}
We investigate possible origins of the trend in financial markets,
where trend we refer to as is a relatively long-term fluctuation 
observed in price change
%	which is a relatively long-term fluctuation observed in price change
(price movement), using a simple deterministic threshold model that
contains no external driving force term to generate trends forcibly.
%	Contrary to common understandings, 
We find that the trend can be generated
by this simple model without any external driving force.
Furthermore, from thorough numerical simulations,
we obtain two following results:
(i)~a trend of monotonic increase or decrease can be generated
only by dealers' minuscule price updates for the next deal trying to follow
an expected forthcoming direction of price change,
(ii)~non-monotonic trends spontaneously emerge when dealers cannot obtain
accurate information about the number of dealers participating in the next deal.
We conclude from these results that the emergence of trends is 
not necessarily generated by an external driving force 
but an natural outcome of the accumulation of minuscule price updates of 
individual dealers with insufficient information about the next deal.
\end{abstract}
%-----------------------------------------------------------------
\begin{keyword}
Price change;
%	Human choice;
%	Incomplete information;
Irregular fluctuations;
Trends
\PACS{02.60.-x, 05.40.-a, 89.65.Gh}
%-------------------------
% 02.60.-x: Numerical approximation and analysis
% 05.40.-a: Fluctuation phenomena, random processes, noise, and Brownian motion
% 89.65.Gh: Economics; econophysics, financial markets, business and management
%-------------------------
\end{keyword}
%-----------------------------------------------------------------
\end{frontmatter}

%%%%%%%%%%%%%%%%%%%%%%%%%%%%%%%%%%%%%%%%%
\section{Introduction}
\label{sec:introduction}
%	Section~\ref{sec:introduction}

\begin{figure}[!t]
\centering
\subfigure[]{\includegraphics[width=6.5cm]{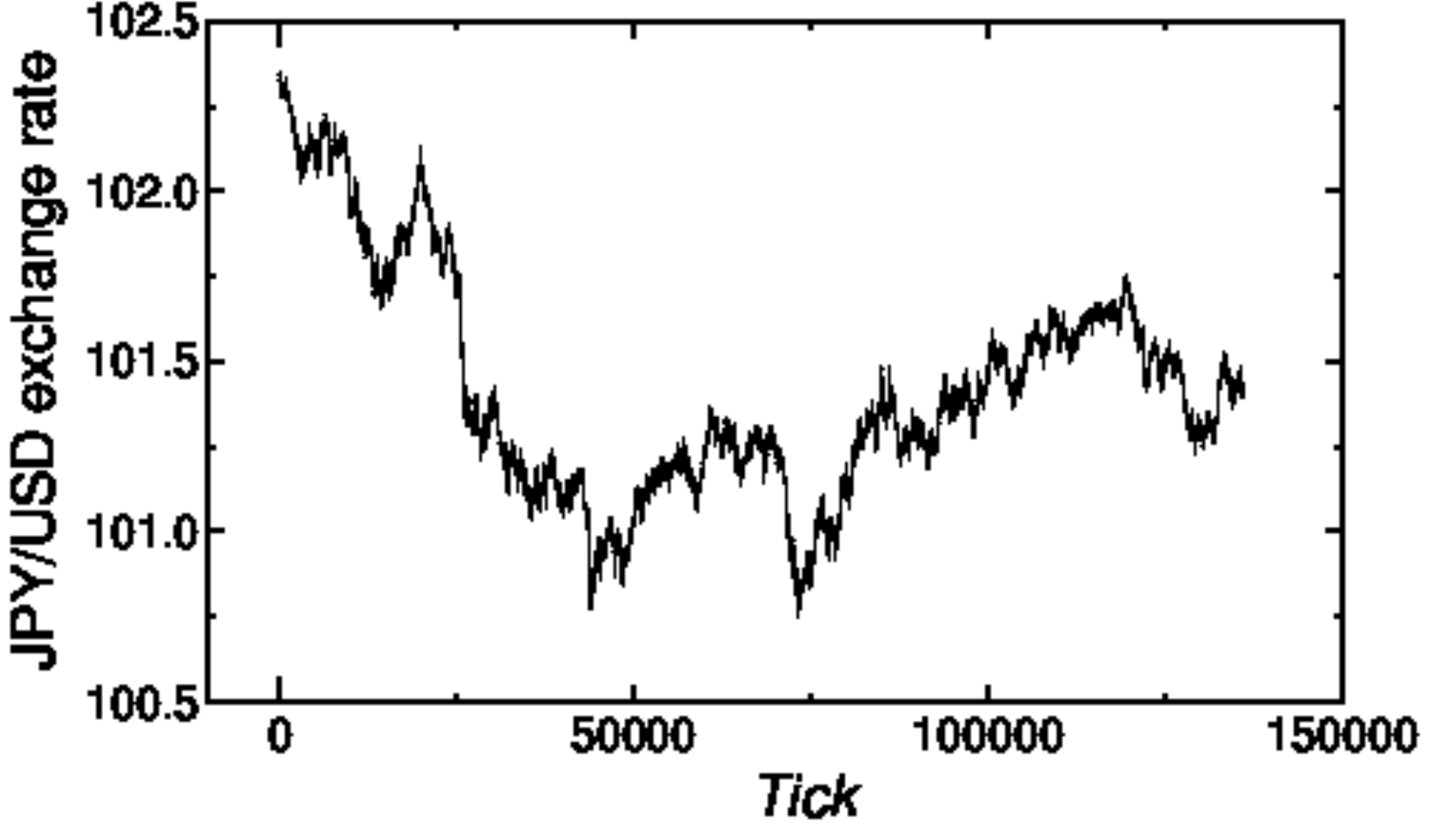}}
~~\subfigure[]{\includegraphics[width=6.5cm]{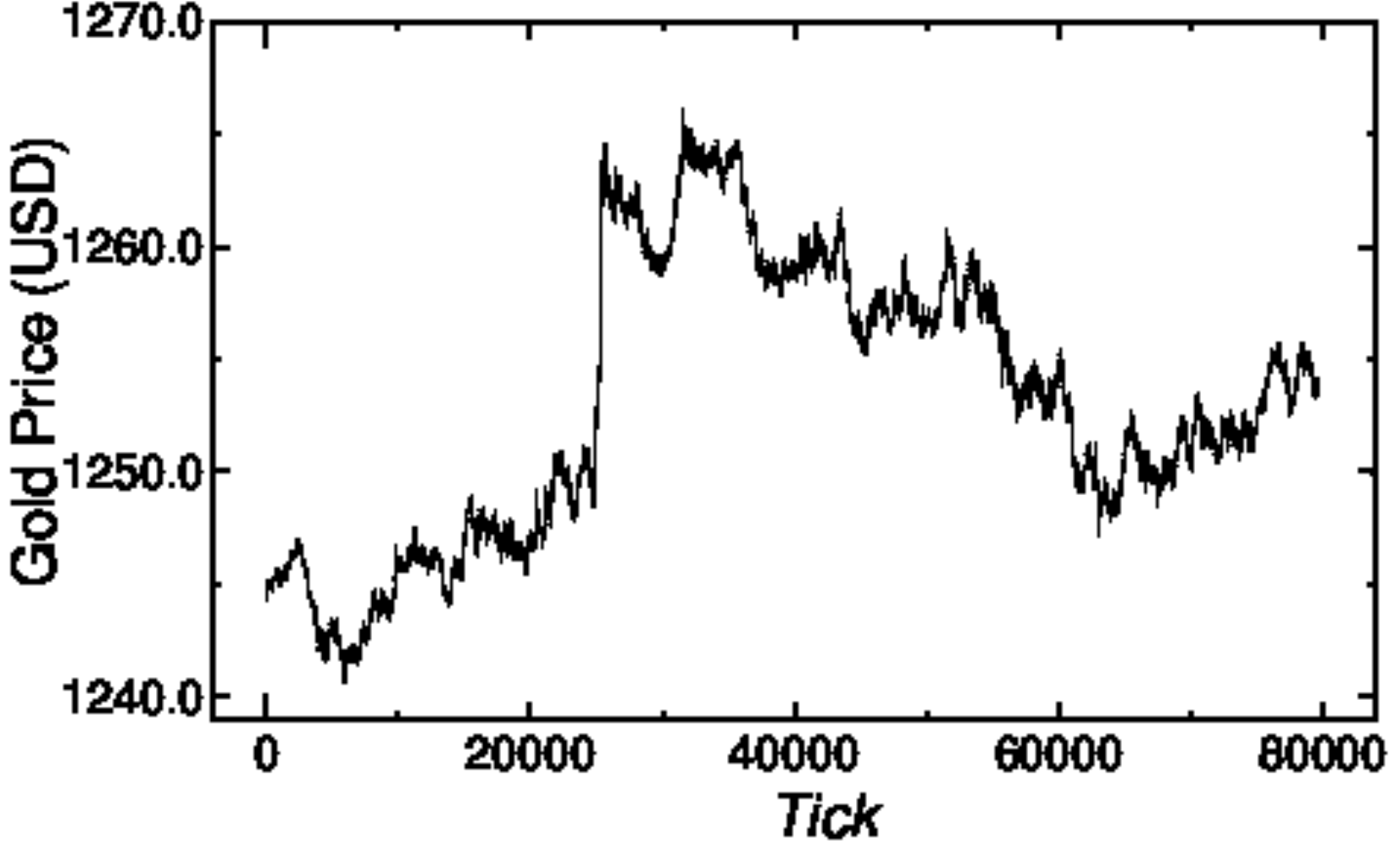}}
\caption{Examples of price changes in financial markets
         showing short-term variabilities and  trends:
         (a)~Tick data of Japanese Yen/US dollar~(JPY/USD) exchange rate 
         and (b)~tick data of gold price
         at New York Mercantile Exchange (NYMEX) futures.
         Both the data are taken for 48 hours from twelve midnight 
         on 3~February~2014 to twelve midnight on 4~February~2014.
         The data can be obtained from \text{http://ratedata.gaincapital.com/}}
    \label{fig:financial_data_with_trends}
%	Fig.~\ref{fig:financial_data_with_trends}
\end{figure}

Many phenomena in human society are dominated by human choices. 
Buying and selling is one of them, 
and we regard financial markets as an aggregation of dealers'
choices and their interactions.
A major indicator in a financial market is its price change~(price movement).
Generally speaking, price change in financial markets fluctuates irregularly,
as shown in Fig.~\ref{fig:financial_data_with_trends}.
However, the mechanism of the fluctuations has not been elucidated yet,
and the question still remains to be challenging.
There are two major approaches to tackle the question.
One is to investigate features or natures of the data
generated by a dynamical system from the viewpoint of 
statistics~\cite{Mantegna-Stanley:nature95,
Ohira-etal:predictability02,Nakamura-Small:RW07}, 
and the other is to construct an artificial market using a model
and investigate events in the market~\cite{Takayasu-etal:dealer_model92,
Sato-Takayasu:dealer_model98,
Ren-etal:minority_games06,
Kaizoji:spin-based_model06,
Maskawa:order-driven_model07,
Alfi-etal:fundamentalist-chartists_model09,
Yamada-etal:dealer_models07,
Yamada-etal:stochastic_dealer_models09}.
The merit of the former approach is that it
provides us with various knowledge and insights
for the understandings of the nature of price changes.
However, this approach is restrictive, 
because it does not always lead us to a deeper understanding of the mechanism 
of price changes produced by the interaction between dealers' choices
and various market price properties~\cite{Yamada-etal:dealer_models07}.
In contrast, the latter approach significantly contributes to clarifying 
the mechanism.
As the purpose of this paper is to find possible 
origins\footnote{We use the term, ``origin,'' as the cause and 
    source of existence.}
of relatively long-term fluctuations generally observed 
in financial market price changes, which we refer to as ``trends,''
we follow the latter approach (we will describe the details of 
our definition of trends later).
%-----------------------------------
%-----------------------------------

Some models, which are categorized as the dealer model,
have been proposed~\cite{Takayasu-etal:dealer_model92,
Sato-Takayasu:dealer_model98,Yamada-etal:stochastic_dealer_models09}.
The dealer model is an agent-based model and constructs an artificial market.
The first dealer model was introduced by Takayasu~{\it et~al.} 
in 1992~\cite{Takayasu-etal:dealer_model92}.
They considered that a market is composed of many dealers
and buying and selling are interactions among them 
with discontinuous~(nonlinear) and irreversible processes.
To implement this mechanism they introduced a numerical model
of financial market prices using threshold dynamics~\cite{Takayasu-etal:dealer_model92}.
In the model a deterministic dynamics is assumed for an assembly
of agents describing mutual trades by threshold dynamics including 
discontinuous irreversible interactions.
After this pioneering work,
numerous studies have been done (for example, 
see Refs.~\cite{Sato-Takayasu:dealer_model98,
Ren-etal:minority_games06,
Kaizoji:spin-based_model06,
Maskawa:order-driven_model07,
Alfi-etal:fundamentalist-chartists_model09,
Yamada-etal:dealer_models07,
Yamada-etal:stochastic_dealer_models09}),
aiming for improvement and refinement to be able to reproduce 
basic empirical laws such as the power-law distribution of price changes,
slow decay of auto-correlation of volatility, and so on.
For the details see~\cite{Yamada-etal:stochastic_dealer_models09}.

\begin{figure}[]
\centering
\subfigure[]{\includegraphics[width=6.5cm]{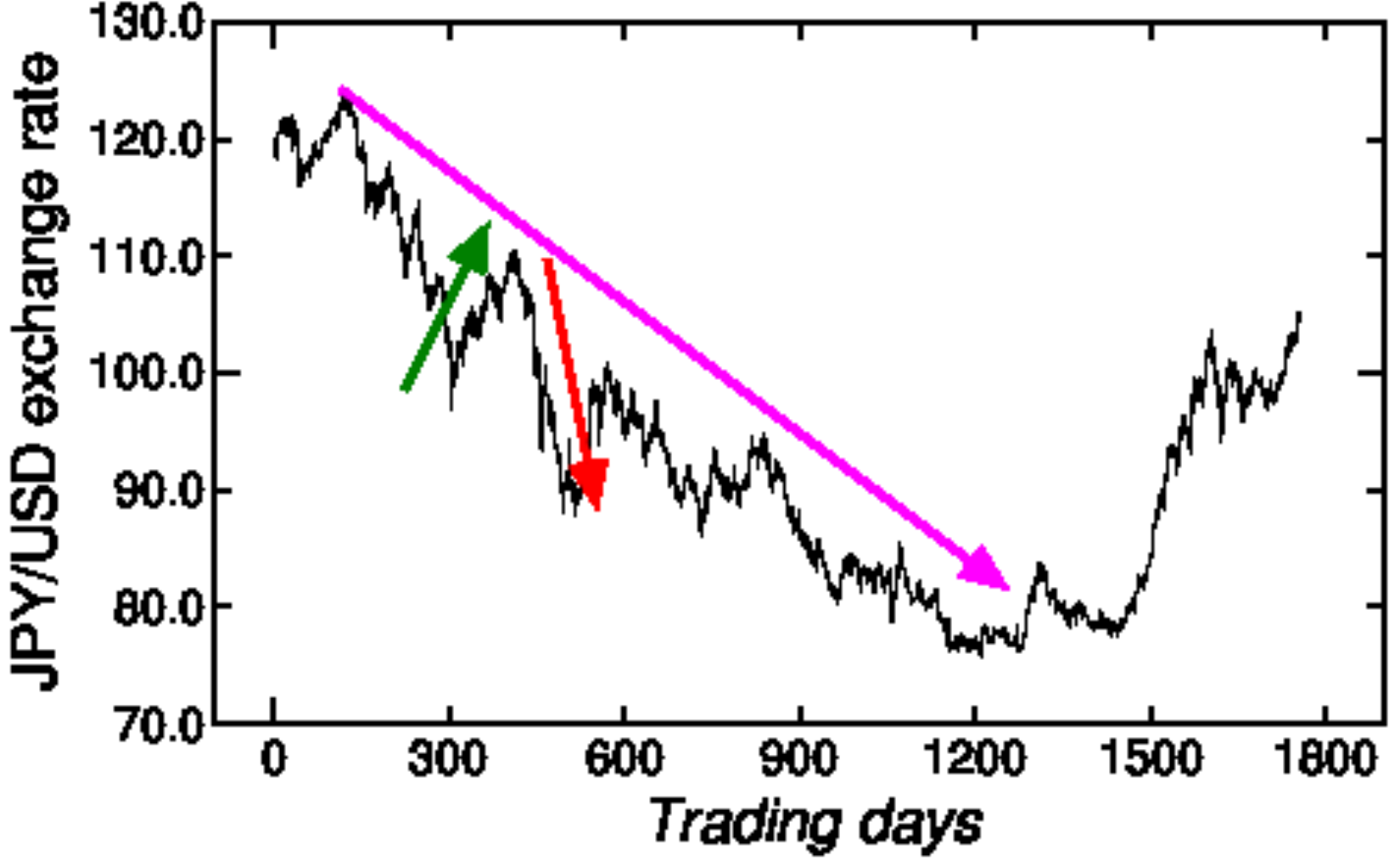}} 
~~\subfigure[]{\includegraphics[width=6.5cm]{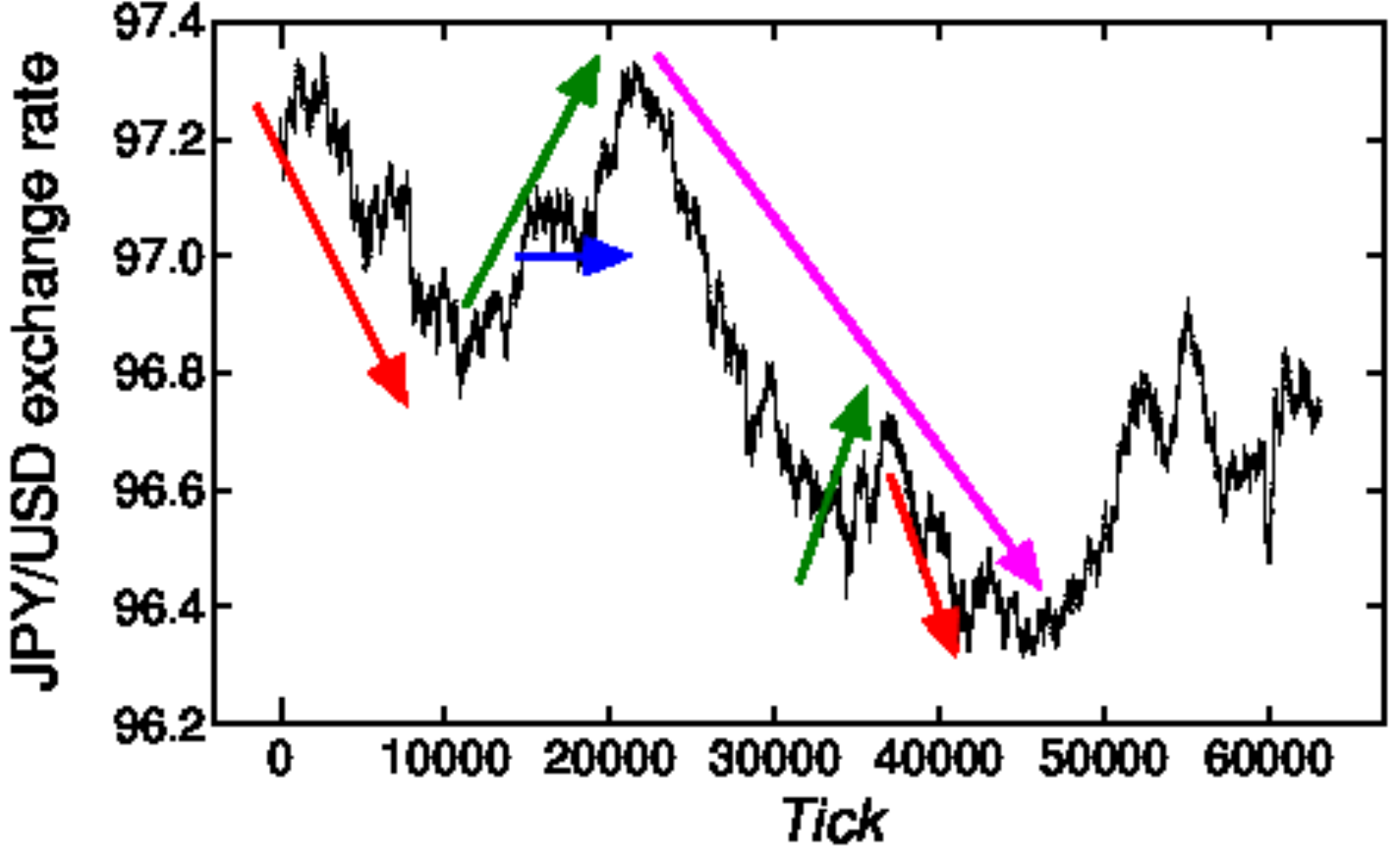}}
\caption{(Colour online) Examples of ``trends'' we recognize: 
         (a)~daily JPY/USD exchange rate from 2 January 2007 to 
         31~December~2013 and (b)~tick data of JPY/USD exchange rate 
         for 24 hours in 7~August~2013 (Wednesday).  
         Although the data are in different resolutions, 
         we can recognize some trends that are marked by the arrows.
         We note that a longer downward trend includes
         shorter upward and downward trends in (a) and (b).}
    \label{fig:trend_example}
%	Fig.~\ref{fig:trend_example}
\end{figure}

In this paper, we observe afresh the behaviours of financial market data carefully.
As mentioned above, price changes in financial markets
show irregular fluctuations (see Fig.~\ref{fig:financial_data_with_trends}).
These irregular fluctuations are usually divided into two main features, 
short-term variabilities and ``trends.''
However, we would like to emphasize that this separation seems to be done
rather arbitrary by simple visual inspection.
A trend we recognize is a general or rough direction of continuous movement
such as upward, downward or sideways during an arbitrary ``long'' period of 
observation compared to the unit time length for taking time series data 
of price change.
If we take a longer observational time,
a trend in a certain period might be considered as a mere fluctuation
around another large trend in the longer period of time 
(See Fig.~\ref{fig:trend_example}).
Whether a movement in price change is recognized as a short-term fluctuation 
or a trend is, therefore, a matter of observational time scale and 
the separation is rather arbitrary.
This is our definition for trends and we recognize trends by this definition.

%-----------------------------------
%-----------------------------------

It should also be noted that there seems to be a vague consensus that
trends are driven by some external factors such as economical 
fundamentals and that short-term variabilities are fluctuations around 
these trends due to stochastic nature of individual deals.
Hence, existing models seems to have been built or improved based on 
this consensus dealing with these two fluctuations (short-term variabilities 
and trends) separately.
A possible origin for the short-term variabilities has already been 
indicated by Takayasu~{\it et~al.}~\cite{Takayasu-etal:dealer_model92}.
Despite of numerous works after the work,
curious to say, discussions on the origins of trends does not seem to be lively,
although market participants are used to pay more attention to the trend 
rather than the details of the short-term variabilities to gain information 
of the overall movement of markets.

One of the major underlying reasons seems to be
that the existence of the trends is taken to be granted
as those driven by some external factors, 
such as economical fundamentals (for example, 
the gross domestic product~(GDP) varying according to policy interest rate, 
remarks by senior administration officials
or central bankers, and so on) and technical analyses of economists.
On the other hand, it is well known that the
trends can also be observed without any remarkable news 
about important economic index changes to be released.
Hence, it brings up a following simple question 
on the origin of trends:
if it were not for these external factors,
do trends completely disappear?
We focus our attention mainly on this point.

The key idea for the present work is the following.
Generally speaking, to understand a certain phenomenon
we need the information about its structure, mechanism, situation, condition 
and so on.
However, it is usually impossible to obtain the entire information.
Although this fact seems to be trivial,
we will show that trends can emerge indeed only from this simple and 
trivial basis.
The conclusion indicates that there is no clear time scale
for dividing the price change clearly into
short-term variabilities and trends.
Both of them spontaneously emerge at the same time.  
In this regard, a trend is not a dynamical movement generated by 
some external driving force but is an accumulation of microscopic price 
updates of individual dealers due to flimsy expectations. 
In this paper, we show two results obtained by our investigation 
based on this idea.
First, we show that a trend of monotonic increase or decrease 
can be generated within the framework of the simplest dealer model 
with no external driving force only by including the effect of dealers' 
minuscule price updates for the next deal trying to follow an expected 
forthcoming direction of the price change.
Next, we show that non-monotonic trends spontaneously emerge 
when dealers cannot obtain accurate information about the number of 
dealers participating in the next deal.
This result implies that trends are spontaneously generated 
even if the influence of fundamentals and technical analyses on 
financial markets is very weak or even absent and 
suggests a possible unknown mechanism for generation of trends.

It should be mentioned here that, 
although we understand that it is important to reproduce 
basic empirical characteristics of real data,
we do not aspire for the reproduction in this paper.
We simply focus on possible origins of trends.\footnote{
    By the progress of computer technology we can obtain 
    high frequency financial data with high resolution time called tick data.
    The term ``tick'' refers to a minute change in the price from deal to deal
    and a set of tick data represents the minimum movement of price change
    produced by financial instruments.
    Various analyses have been done using the tick data.
    However, we note that all deals are not precisely recorded in a set of tick data.
    Tick data we usually obtain are measured at certain time intervals.
    There are not many discussions about discrepancy of statistical nature
    between intermittent data and data with all deals.
    In this paper, we are not going deep into the analysis of real data.}

This paper is organized as follows. 
We start from an existing deterministic dealer model with threshold elements
and add some essential modifications to the original dealer model
based on the key idea of dealers' incomplete information acquisition of the next deal.
In Section~\ref{sec:current_technologies},
we briefly review the original dealer model and its basic behaviours.
In Section~\ref{sec:exploration_of_origins_for_trends}
we extend the original dealer model to include microscopic and distinctive price
updates of individual dealers and show that trends are generated
only by this extension without introducing any external driving force term.
Section~\ref{sec:discussion&summary} is for the discussions and summary.

%%%%%%%%%%%%%%%%%%%%%%%%%%%%%%%%%%%%%%%%
\section{Current technologies: deterministic threshold dealer model}
\label{sec:current_technologies}
%	Section~\ref{sec:current_technologies}

To generate trends and find possible (or unknown) origins of trends, 
we use a previously proposed deterministic dealer model with threshold elements.
As the original model is proposed in 1992
by Takayasu {\it et~al.}~\cite{Takayasu-etal:dealer_model92},
we refer to the model as the ``dealer model~92.''
Since the dealer model~92 is deterministic including no probabilistic factor,
we expect that we are able to have a better sense of the behaviours generated 
by the model.
As the model treats price change of all deals,
the price data generated by the model should be compared to tick data.
For simplicity the deals in the model are assumed to be concerning to only 
one brand\footnote{
The precise meaning of ``brand'' is an individual financial commodity.
}
in one market.
Also, we assume that all dealers participate deals with the same rule.

After the dealer model~92 was proposed, some modifications have been 
made~\cite{Sato-Takayasu:dealer_model98,Yamada-etal:dealer_models07,
Yamada-etal:stochastic_dealer_models09}.
However, we consider that the modifications are artificial, 
since all deals are done by only two dealers, that is, one buyer and one seller.
In the actual deals there are usually more than one sellers,
which we consider is one of the vital aspects of deals.
As the dealer model~92 treats such a situation, 
we adopt the dealer model~92.\footnote{The dealer model~92 was modified 
    by Sato and Takayasu in 1998~\cite{Sato-Takayasu:dealer_model98}.
    We refer to the modified model as the dealer model~98.
    The dealer model~98 has the opposite assumption to the dealer model~92.
    The dealer model~98 assumes that all dealers have a small amount of properties
    and basically change their attitudes (that is, a buyer becomes 
    a seller and a seller becomes a buyer in the next deal)
    when the deals are done.
    Unlike the dealer model~92, the dealer model~98 is tractable and 
    can generate irregular fluctuations over long periods 
    like Fig.~\ref{fig:financial_data_with_trends}.
    This seems to be one of the supposed reasons
    that the subsequent market price analyses are developed 
    based on the dealer model~98~\cite{Yamada-etal:dealer_models07,
    Yamada-etal:stochastic_dealer_models09}.}

In this section we briefly review the dealer model~92
and clarify the mechanism of the price change.

%%%%%%%%%%%%%%%%%%%%%%%%%%%%%%%%%%%%%%%%
\subsection{Dealer model~92}
\label{subsec:dealer_model92}
%	Section~\ref{subsec:dealer_model92}

Deals, composed of consecutive selling and buying, form
a typical discontinuous and irreversible process.
A deal is done, if a buying condition
and a selling condition meet,
while it is undone, if not.
This is the origin of the discontinuity.
The buyer and the seller of the last deal would never deal again
under the identical condition at least for a while.
This is the origin of the irreversibility.
The dealer model~92 is introduced to reflect these aspects of dealing
and we here begin by following the basic settings of the original 
model~\cite{Takayasu-etal:dealer_model92}.

The market is composed of $ N $~dealers.
Let the $ i $-th dealer's buying price and selling price be 
$ B_i $ and $ S_i $, respectively.\footnote{Buying price~$ B_i $ and 
    selling price~$ S_i $ are called Ask and Bid in financial markets, 
    respectively.}
The selling price~$ S_i $ is larger than the buying price~$ B_i $, 
and the difference~$ L_i = S_i - B_i $ is always positive.
For simplicity we use a constant value~$ L $ for all
$ L_i $~\cite{Takayasu-etal:dealer_model92}.\footnote{
We use a constant value for all $ L_i $.  
Although one might think that it is too simple to be realistic,
one of the purposes of this paper is to show that trends appear indeed
even under this simplest setting.
}
This simplification gives $ L = S_i - B_i $ for all $ i $. 
A deal is done between pairs of $ i $ and $ j $ when $ B_i \geq S_j $.
In this model, a deal among all dealers is done 
when the following condition is satisfied:
%-----------------------------------
\begin{equation}
    \max \{ B_i \} - \min \{ B_i \} \geq L,
    \label{eq:deal_condition}
%	Eq.~(\ref{eq:deal_condition})
\end{equation}
%-----------------------------------
where $ \max \{ B_i \} $ and $ \min \{ B_i \} $ indicate the maximum 
and minimum values.\footnote{Equation (\ref{eq:deal_condition}) has 
    the same meaning of $ \max \{ B_i\} - \min \{ S_i\} \geq 0 $
    because $ S_i = B_i + L $.}
In the dealer model~92 there is one buyer and one or more than one sellers.
The buyer~$ i $ is the dealer who gives the highest buying price $ B_i $
and can buy at the price $ B_i $ from other dealers~$ j $ who satisfy the following selling condition
%-----------------------------------
\begin{equation*}
    \max \{ B_i\} - \{ B_j \} \geq L.
\end{equation*}
%-----------------------------------
The price~$ P(t) $ of the deal is therefore defined by the $ \max \{ B_i\} $.
If Eq.~(\ref{eq:deal_condition}) is not satisfied,
the deal is not done and the market price does not change.
Hence,
%-----------------------------------
\begin{eqnarray}
    P(t) = 
    \begin{cases}
    \max \{ B_i\} & \text{(when a deal is done)}, \\
    P(t-1) & \text{(when a deal is not done)}.
    \end{cases}
    \label{eq:price92}
%	Eq.~(\ref{eq:price92})
\end{eqnarray}
%-----------------------------------

When a deal is done,
there are one buyer, sellers who could make the deal,
and the remainders who could not participate in the deal.
However, the remainder do not look wistfully 
and enviously at the deal with doing nothing.
As all dealers are potentially willing to participate in the deals,
they re-establish their prices for the next deal.
Obviously, dealers who want to buy need to offer higher price
and dealers who want to sell need to offer cheaper price.
To reflect the eagerness to participate in the next deal,
the $ i $-th dealer's expectation~$ a_i $ is introduced.
The term represent a character of the $ i $-th dealer.
If $ a_i > 0 $, the dealer raises the price setting,
and if $ a_i < 0 $, the dealer lowers the price setting.
That is, dealers with $ a_i > 0 $ and $ a_i < 0 $ 
take the actions as buyers and sellers, respectively.

In this model each dealer is assumed to have his/her own expectation.
Hence, the values of $ a_i $ are chosen from uniform random numbers
in a range and the mean of $ \{ a_i \} $ is set to be zero, 
which keeps the overall expectations of dealers
for buying and selling in equilibrium.\footnote{Broadly speaking, 
    there are three economic expectations: 
    (i)~myopic expectation, (ii)~adaptive expectation 
    and (iii)~rational expectation.
    In the actual financial markets a dealer's expectation~$ a_i $ 
    should be given based on some reasons.
    However, we do not overinterpret it in this paper.}
As dealers becomes buyers and sellers depending on their conditions 
and circumstances in the actual financial markets,
the details inside of the deals are very complicated.
To simplify the situation, however, as the model assumes that all dealers have 
infinitely large amount of property, 
the dealers do not change their attitudes even if a deal is done
(that is, the sign of $ a_i $ does not change)~\cite{Takayasu-etal:dealer_model92,Sato-Takayasu:dealer_model98}.
Also, this model is assumed to be in an invariant state (steady state).
Hence, the values of $ a_i $ do not change once 
assigned~\cite{Takayasu-etal:dealer_model92}.

An unique aspect of this model is that
it takes into account an acquisitive nature of the buyer and 
sellers for the next deal.
It is assumed that the buyer and sellers who could
participate a deal expect that
they can participate again in the next deal with a cheaper price for the buyer 
and a higher price for the sellers.
The other dealers who could not participate in the previous deal
do not have such a psychological tendency.
To reflect such a tendency,
the term~$ \Delta_i $ is introduced:
%-----------------------------------
\begin{eqnarray}
    \Delta_i = 
    \begin{cases}
    -\delta & \text{(for the buyer)}, \\
    \delta/n & \text{(for the sellers)}, \\
    0 & \text{(for the nonparticipants of the deal)},
    \end{cases}
    \label{eq:Delta}
%	Eq.~(\ref{eq:Delta})
\end{eqnarray}
%-----------------------------------
where $ 0 < \delta < L $ and $ n $ is the number of the sellers
of the deal.
That is, the next buying price of the buyer falls
by the amount of $ \delta $ from the current buying price and the next selling prices of 
the sellers rises by the amount of $ \delta/n $ from the current selling prices.
The buyer and sellers who could not participate in the next deal have no anticipation.
Also, when Eq.~(\ref{eq:deal_condition}) is not satisfied 
(that is, a deal is not done), $ \Delta_i = 0 $ for all $ i $.
Note that the summation of all $\Delta_i$ is zero.
%-----------------------------------
\begin{figure}[!t]
\centering
\includegraphics[width=10.0cm]{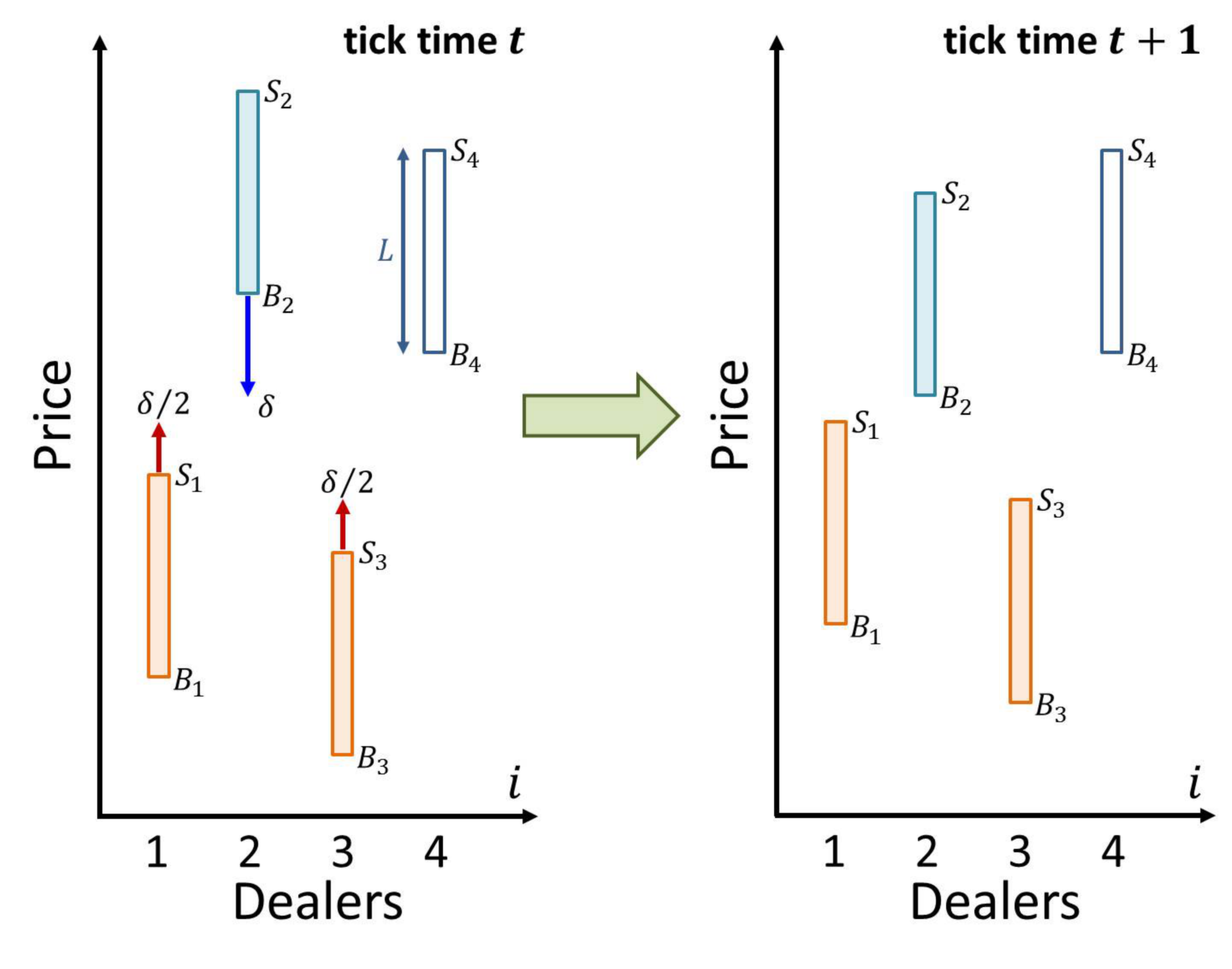}
\caption{(Colour online)~Schematic diagram of $ \Delta $
    from time $ t $ to $ t+1 $ when Eq.~(\ref{eq:deal_condition}) is satisfied.
    The length of each vertical bar is the constant~$ L $.
    The top and bottom of each bar correspond to the upper and lower 
    threshold price of each dealer, respectively.
    The $ \Delta $ of the buyer is $ -\delta $ and that of the sellers
    is $ \delta/n $, where $ 0 < \delta < L $ and $ n $ is the number of 
    the sellers.
    The value of $ \delta $ is constant defined by Eq.~(\ref{eq:Delta}).
    In this figure the buyer is $ i=2 $, 
    the sellers are $ i=1 $ and $ 3 $~($ n=2 $),
    and the dealer~$ i=4 $ does not participate in the deal.}
    \label{fig:delta}
%	Fig.~\ref{fig:delta}
\end{figure}
%-----------------------------------

Based on these ideas, the price update for the $ i $-th dealer
in the dealer model~92 is given by
%-----------------------------------
\begin{equation}
B_i(t+1) = B_i(t) + \Delta_i(t) + a_i + c_i \{ P(t) - P(t_\text{prev}) \},
    \label{eq:org_dealer_model92}
%	Eq.~(\ref{eq:org_dealer_model92})
\end{equation}
%-----------------------------------
where the term~$ c_i $ characterizes the $ i $-th dealer's response to 
the change of market price and $ t_\text{prev} $ indicates the time 
when the last deal is done~\cite{Takayasu-etal:dealer_model92}.
The behaviour of Eq.~(\ref{eq:org_dealer_model92}) has already been scrutinized for various
values of $ c_i $, $ a_i $ and the number of dealers, $ n $ and
it is commonly known that Eq.~(\ref{eq:org_dealer_model92}) is extremely unstable 
and uncontrollable when $ c_i~\neq~0 $~\cite{Takayasu-etal:dealer_model92,
Takayasu-etal:dealer_model93}.
For details see~\cite{Takayasu-etal:dealer_model93}.
Hence, we use Eq.~(\ref{eq:org_dealer_model92}) with $ c_i = 0 $ as the dealer model~92
for the starting point of our work, which becomes
%-----------------------------------
\begin{equation}
B_i(t+1) = B_i(t) + \Delta_i(t) + a_i.
    \label{eq:dealer_model92}
%	Eq.~(\ref{eq:dealer_model92})
\end{equation}
%-----------------------------------

%%%%%%%%%%%%%%%%%%%%%%%%%%%%%%%%%%%%%%%%
\subsection{Typical behaviours of the dealer model~92}
\label{subsec:typical_behaviours}
%	Section~\ref{subsec:typical_behaviours}

We show typical behaviours generated by the dealer model~92 with Eq.~(\ref{eq:Delta})
for $\Delta_i$.
To generate the data we take the number of dealers $ N = 100 $
and $ \delta = 0.4 $.
The initial value of $ B_i(t) $ is taken from uniform random numbers 
in the range $ (-L, L) $ with $ L = 1 $.
The values of $ a_i $ in Eq.~(\ref{eq:dealer_model92}) 
are taken from uniform random numbers in the range~$ (-\alpha, \alpha) $ 
with $ \alpha = 0.01 $.
As mentioned above, to equalize dealers' eagerness for buying and selling, 
the mean of $ \{ a_i \} $ in Eq.~(\ref{eq:dealer_model92}) 
is set to be zero.\footnote{This setting for the mean of $ \{ a_i \} $ 
    is not mentioned in~\cite{Takayasu-etal:dealer_model92}
    and is mentioned very casually in~\cite{Takayasu-etal:dealer_model93}.
    However, we note that the setting is crucial. 
    Without this setting, the data generated by 
    the dealer model~92 with Eq.~(\ref{eq:Delta}) only show monotonic increase or 
    decrease in almost all cases,
    even if we use Eq.~(\ref{eq:dealer_model92})
    (that is, Eq.~(\ref{eq:org_dealer_model92}) with $ c_i = 0 $).}
%-----------------------------------
\begin{figure}[!t]
\centering
\subfigure[]{\includegraphics[width=6.5cm]{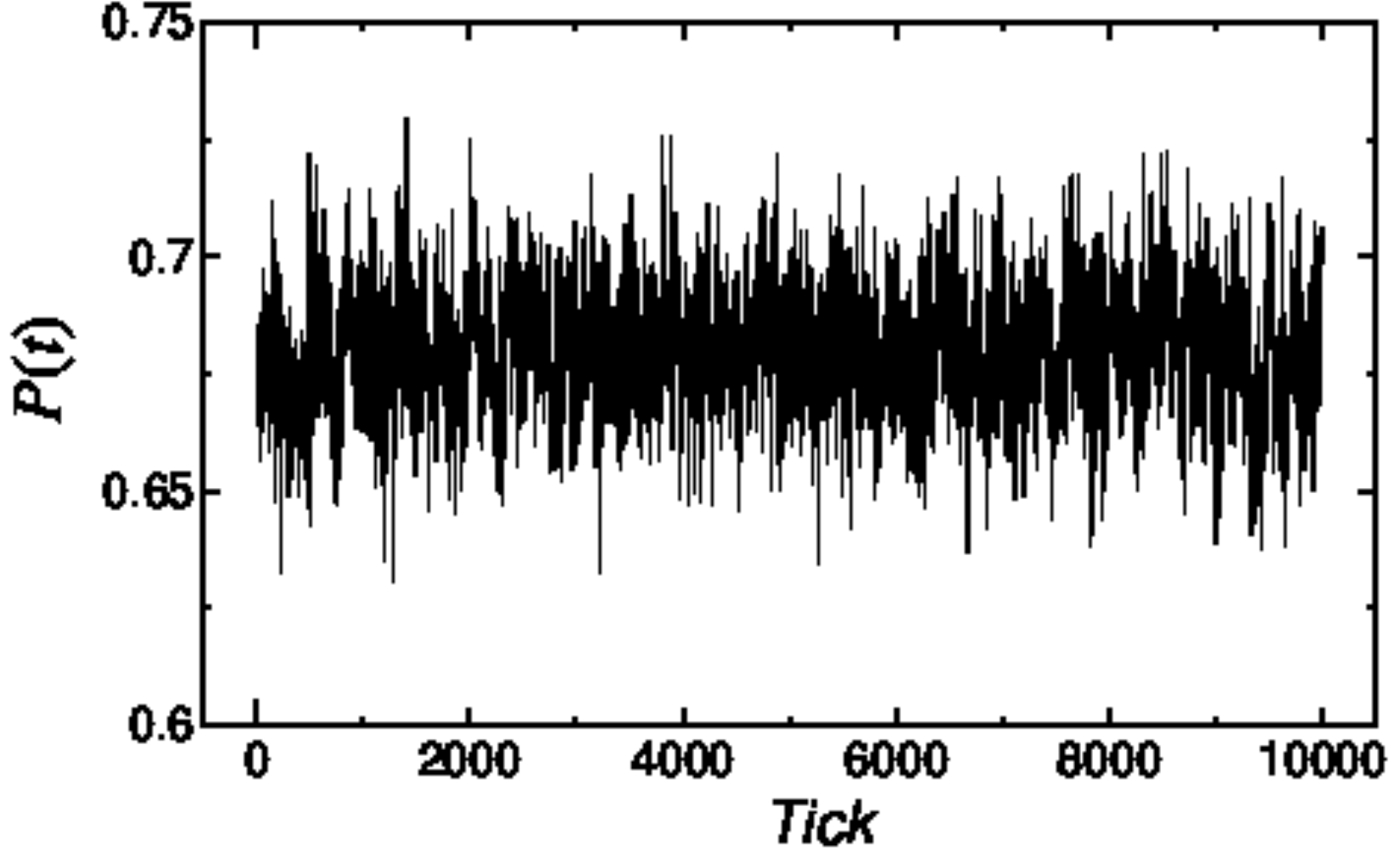}} 
~~\subfigure[]{\includegraphics[width=6.5cm]{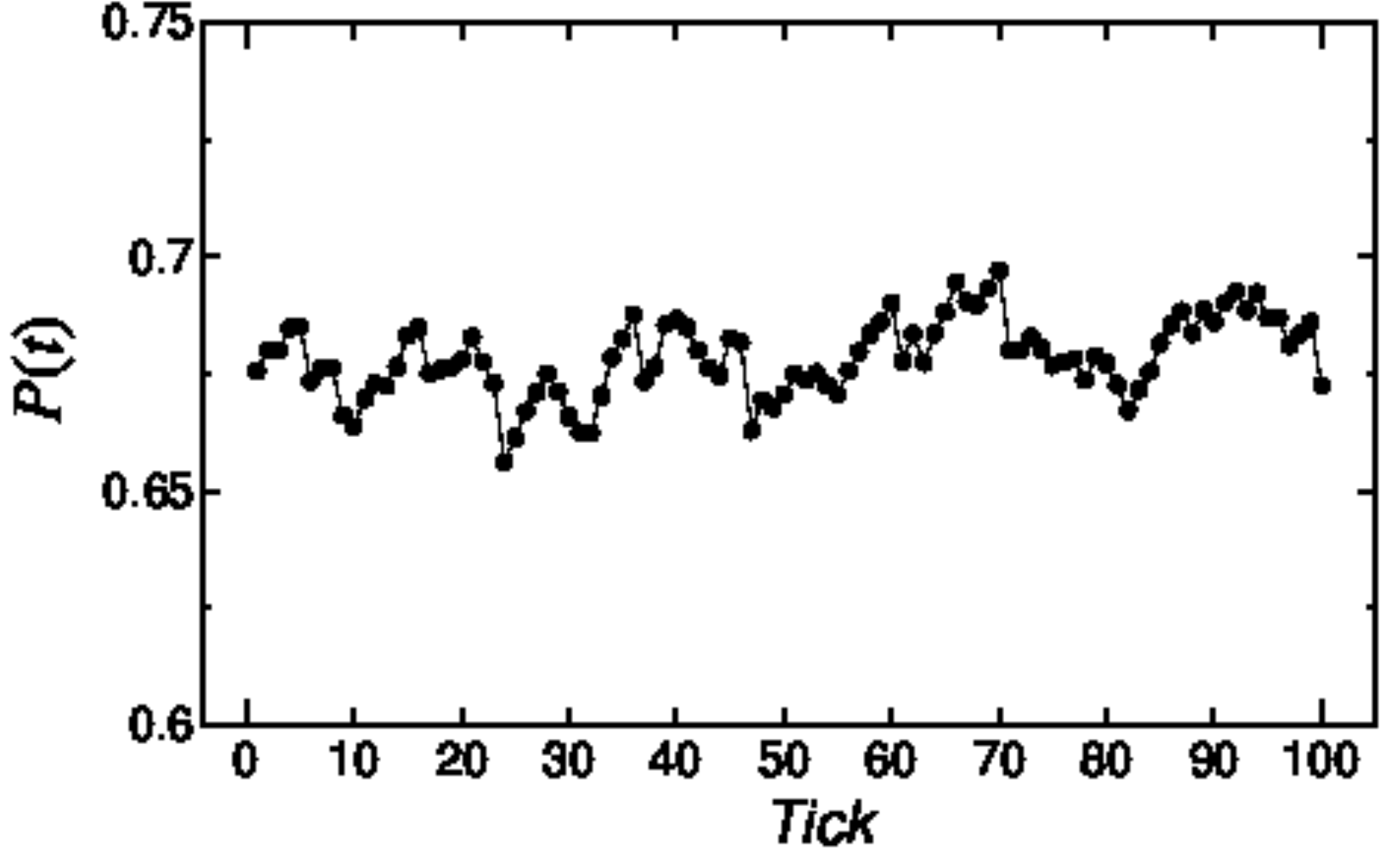}}
\caption{Typical behaviours of simulated price data~$ P(t) $ 
         of the dealer model~92, Eq.~(\ref{eq:dealer_model92}),
         with Eq.~(\ref{eq:Delta}),
         where (b) is an enlargement of (a), 
         and both of black circle mark and line are used to show details of 
         price changes in (b).
         For the model $ N = 100 $, $ L = 1 $, $ \alpha = 0.01 $ 
         and $ \delta = 0.4 $ are used.
         The value of $ a_i $ and the initial condition of $ B_i(t) $ 
         are taken from uniform random numbers 
         in the ranges~$ (-\alpha, \alpha) $ and $ (-L, L) $,
         respectively.}
    \label{fig:behaviours_of_models92}
%	Fig.~\ref{fig:behaviours_of_models92}
\end{figure}
%-----------------------------------

Figure~\ref{fig:behaviours_of_models92} shows the typical behaviours
of price data~$ P(t) $, where we use the price data only when the deal is done.
In Fig.~\ref{fig:behaviours_of_models92}(a), it is shown that
the behaviours are stable over long periods and $ P(t) $ fluctuates 
around the mean value showing short-term variabilities but does not have trends.
Figure~\ref{fig:behaviours_of_models92}(b) shows that
the price~$ P(t) $ does not change drastically but slightly over time.

%%%%%%%%%%%%%%%%%%%%%%%%%%%%%%%%%%%%%%%%
\subsection{Some observations and possible reasons for lacking of trends}
\label{subsec:some_observations&reasons_for_no_trend}
%	Section~\ref{subsec:some_observations&reasons_for_no_trend}

The zero value of the mean of the dealer's expectation~$ \{ a_i \} $
means that the overall summation of the dealer's expectation is zero.
We note that this does not necessarily mean that the number of dealers with $ a_i > 0 $ 
should be the same as that of dealers with $ a_i < 0 $.
The number of the dealers with $ a_i > 0 $ for Fig.~\ref{fig:behaviours_of_models92} 
is 38 where the number of the total dealers is 100.\footnote{This number 
    of the buyers is slightly extreme examples.  
    As $ \{ a_i \} $ is uniformly distributed between $ -\alpha $ and $ \alpha $ 
    with $ \alpha > 0 $,
    the number with $ a_i > 0 $ is usually around half of the total.}
From the observations,  trends do not appear 
when dealers' expectations for buying and selling equal out.
In other words, breaking the balance of dealers' expectation of buying and selling
could be vital for generating trends.
Based on this idea, we make attempts to generate trends
under some well-defined situations in the next section.

%%%%%%%%%%%%%%%%%%%%%%%%%%%%%%%%%%%%%%%%
\section{Exploration of origins for trends}
\label{sec:exploration_of_origins_for_trends}
%	Section~\ref{sec:exploration_of_origins_for_trends}

In this section we make attempts to generate trends by extending 
the dealer model~92 to the directions suitable to two realistic situations.
The one is the trend with supposition.  
We show that a trend emerges from the intention of individual dealers 
following to the expected direction of forthcoming price change.
The other is the trend with no supposition. 
We show that a trend emerges almost spontaneously only from incomplete information
of the dealers even if they do not have any intention for the deals.
After considering these two trends separately, we combine both ideas into one model
and generate data using it.

%%%%%%%%%%%%%%%%%%%%%%%%%%%%%%%%%%%%%%%%
\subsection{Trends with supposition: premeditated trends}
\label{subsec:trends_with_supposition}
%	Section~\ref{subsec:trends_with_supposition}

We show here that we are able to generate monotonic increase or decrease 
trends at our disposal only by introducing the effect of intentions of 
individual dealers following to the expected direction of forthcoming 
price change.
It is widely believed that continuous trends 
in relatively long-terms
are generated by fundamentals.
It is considered that fundamentals influence dealers' longer-term intention 
and the major trends in the relative long term are believed to be 
generated by the dealers' supposition.
We consider that the effect of fundamentals comes into the dealer model
through the supposition of dealers for the direction of the price change 
in a relatively long term due to the changes in fundamentals.\footnote{
We note that the correspondence between the types of fundamentals 
and the types of trends is basically unknown.  
Hence, we therefore avoid the overinterpretation of this relationship and 
do not touch on the problem.  
We simply focus on the problem of how we can generate monotonic increase 
and decrease trends by dealers' supposition.}

As Eqs.~(\ref{eq:Delta}) and~(\ref{eq:dealer_model92}) show, 
when a deal is done, the buyer brings down
the asking price for the next deal,\footnote{Although
    the price brought by $ a_i $ goes up and that by $ \delta $ goes down, 
    the total asking price falls as a result, as $ \delta $ is larger than $ a_i $.}
expecting that it might be possible to buy at a cheaper price.
Similarly, when a deal is done, the sellers bring up
their selling prices for the next deal,
expecting that it might be possible to sell at a higher price.

How does the acquisitiveness change in the existence of
an expected direction of price change in a market?
When an increase (or decrease) direction of price change is expected in a market,
% which corresponds to self-fulfilling expectation,
it seems to be plausible that
a dealer, whether the dealer is either a buyer or a seller, considers that
other dealers are also not likely to take the same attitude in the next deal
if the dealer offers a higher (or cheaper) price.
To incorporate this idea, we redefine Eq.~(\ref{eq:Delta}) with the small values
$ \varepsilon_{b} $ and $ \varepsilon_{s} $ as follows:
%-----------------------------------
\begin{eqnarray}
    \Delta_i = 
    \begin{cases}
    -\delta \times (1.0 + \varepsilon_{b}) & \text{(for the buyer)}, \\
    (\delta \times \varepsilon_s)/n & \text{(for the sellers)}, \\
    0 & \text{(for the nonparticipants of the deal)},
    \end{cases}
    \label{eq:Delta_with_epsilon}
%	Eq.~(\ref{eq:Delta_with_epsilon})
\end{eqnarray}
%-----------------------------------
where $ 0 < \delta < L $ and $ n $ is the number of the sellers of the deal
as taken in Eq.~(\ref{eq:Delta}).
For newly introduced parameters, $ \varepsilon_b $ and $ \varepsilon_s $,
$ -1 \leq \varepsilon_{b} \leq 0 $ and $ \varepsilon_{s} \geq 0 $
when a monotonic increase of price change is expected,
and $ 0 \leq \varepsilon_{b} \leq 1 $ and $ \varepsilon_{s} \leq 0 $
when a monotonic decrease of price change is expected.

%-----------------------------------
\begin{figure}[!t]
\centering
\subfigure[]{\includegraphics[width=6.5cm]{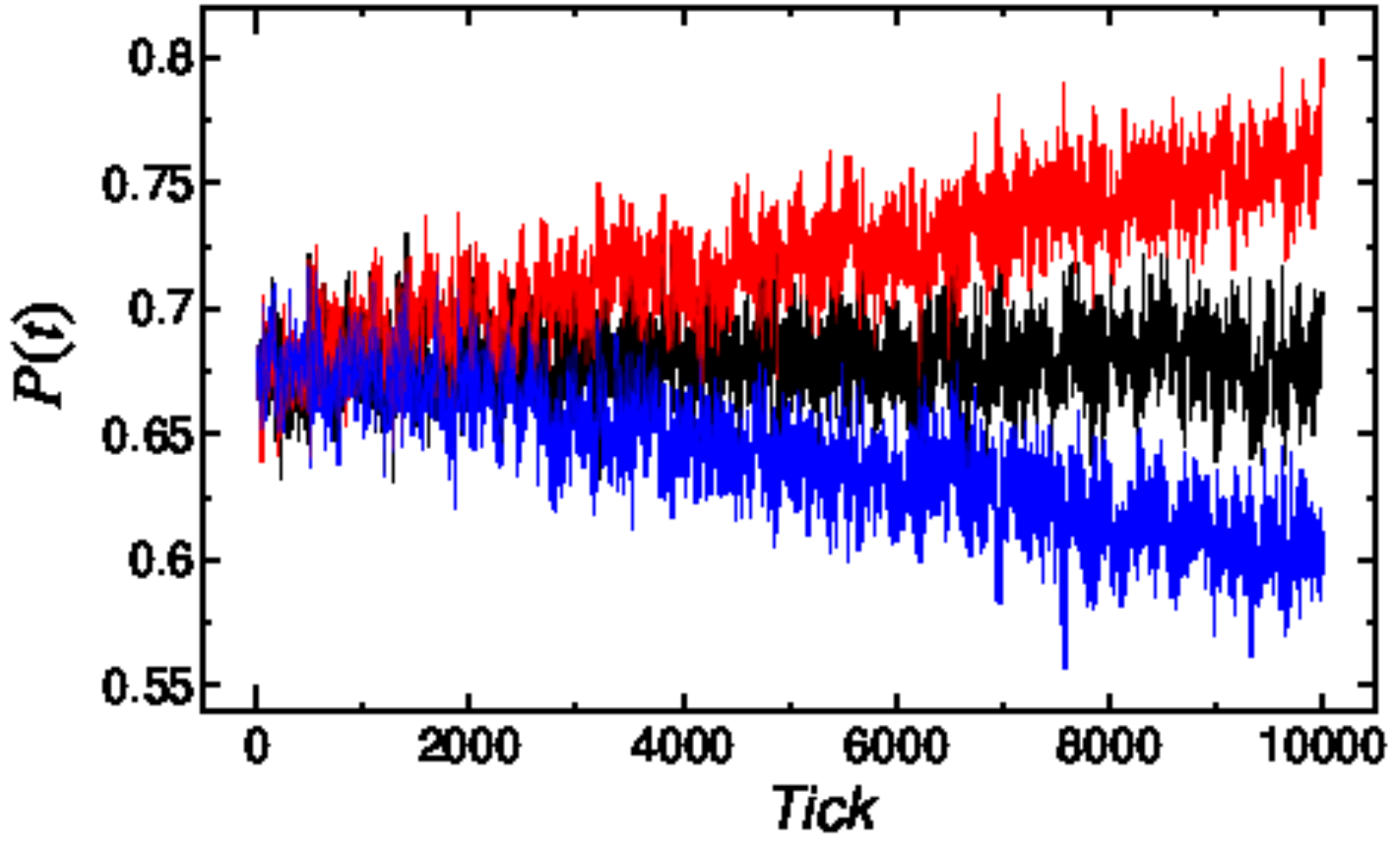}} 
~~\subfigure[]{\includegraphics[width=6.5cm]{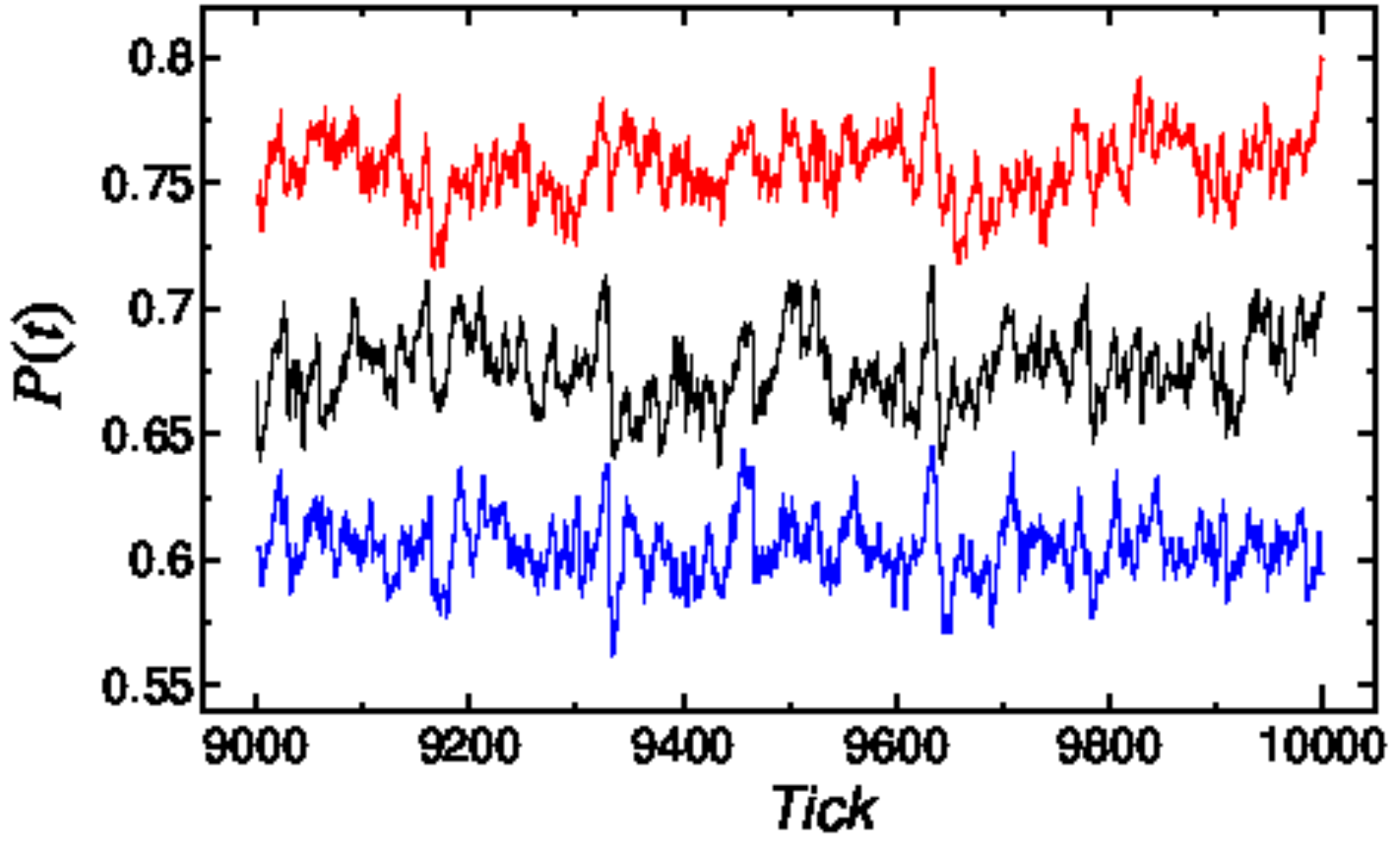}}
\caption{(Colour online)~Monotonic increase and decrease trends generated 
         by dealer model~92 with Eq.~(\ref{eq:Delta_with_epsilon}), 
         where (b) is an enlargement of (a).
         $ \varepsilon_b = -0.002 $ for a monotonic increase trend,
         $ \varepsilon_b = 0.002  $ for a monotonic decrease trend,
         and $ \varepsilon_{s} = 0 $ for both the cases.
         The data with no trend is the same as that shown 
         in Fig.~\ref{fig:behaviours_of_models92}(a).
         We show it for the for comparison.
         All other conditions are the same in Fig.~\ref{fig:behaviours_of_models92}.}
    \label{fig:monotonic_change92}
%	Fig.~\ref{fig:monotonic_change92}
\end{figure}
%-----------------------------------
Figure~\ref{fig:monotonic_change92} is the data generated 
by the dealer model~92 with Eq.~(\ref{eq:Delta_with_epsilon})
instead of Eq.~(\ref{eq:Delta})
where $ \varepsilon_b = \pm 0.002 $ and $ \varepsilon_{s} = 0 $.
We find a monotonically increasing trend for $ \varepsilon_b = -0.002 $
and a monotonically decreasing trend for $ \varepsilon_b = 0.002 $
on Fig.~\ref{fig:monotonic_change92}(a).
It should be noted that,
as these monotonically increasing and decreasing trends are very slow, 
we cannot recognize the trends clearly if we take shorter period of time for observation,
as is shown in Fig.~\ref{fig:monotonic_change92}(b).
We have also confirmed that the dealer model~92 with
$ (\varepsilon_b, \varepsilon_s) = (0, 0.002) $ show similar behaviours.
Although the values of the parameters are small,
this tiny little ulterior motive generates major trends.
The results indicate that
the trends are built on a delicate balance of acquisitive natures 
between the buyers and sellers.
The results also indicate that,
even if the amount of a change in the psychological tendency of a dealer
is very small, it can bring a major influence on the market prices.

When we use larger values of $ \varepsilon_b $ or $ \varepsilon_s $
or when we use the pair of $ \varepsilon_b $ and $ \varepsilon_s $ together,
we have confirmed to obtain steeper trends.

%%%%%%%%%%%%%%%%%%%%%%%%%%%%%%%%%%%%%%%%
\subsection{Trends with no supposition: unpremeditated trends}
\label{subsec:trends_with_no_supposition}
%	Section~\ref{subsec:trends_with_no_supposition}

In the previous section we have confirmed that monotonic trends emerge by 
dealers' expectation of the forthcoming price change,
even if it is minuscule.
In this case we can guess or estimate the underlying expectation of
individual dealers based on their behaviours.
In numerical calculations,
we can obtain complete information about all deals.
It should be emphasized, however, that it is almost impossible
to acquire complete information in actual deals.
In this section, we focus on the case in which only incomplete
information on the deals is available and the influence of fundamentals 
and technical analyses on financial markets are almost absent.

Let us carefully examine an important feature of the dealer model~92.
Generally speaking, it is impossible that all dealers in actual financial markets
exactly know the number of sellers in every deal in advance.
% when a deal is done.
However, the number of sellers~$ n $ is included in Eq.~(\ref{eq:Delta})
to reflect the acquisitive nature of the dealers.
That is, the dealer model~92 contains a parameter which is impossible to know 
in advance in actual deals, though
this parameter~$ n $ plays a crucial role in the model.
As Eq.~(\ref{eq:Delta}) shows,
the buyer's price falls $ \delta $
and each of $ n $ sellers' prices rises $ \delta/n $
when a deal is done.
In other words,
the total amount of the rises of the sellers' prices~$ n \times \delta/n $
exactly compensates for the decline of the buyer's price~$ \delta $.
Because of this exact compensation effect,
the price data generated by the dealer model~92 with Eq.~(\ref{eq:Delta})
have only short-term variabilities and do not have trends.

\begin{figure}[!t]
\centering
\includegraphics[width=6.5cm]{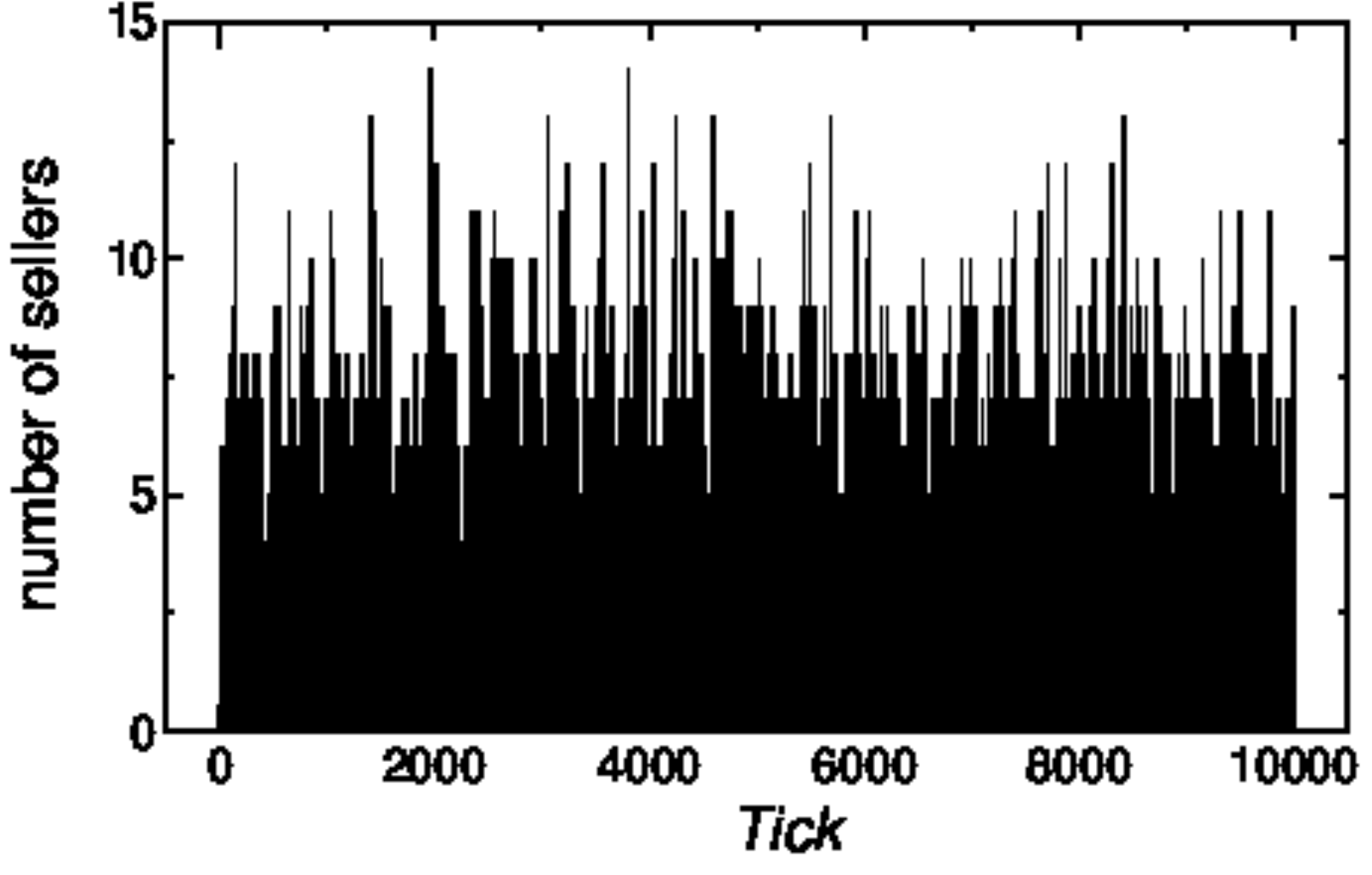} \\
(a) \\
\includegraphics[width=6.5cm]{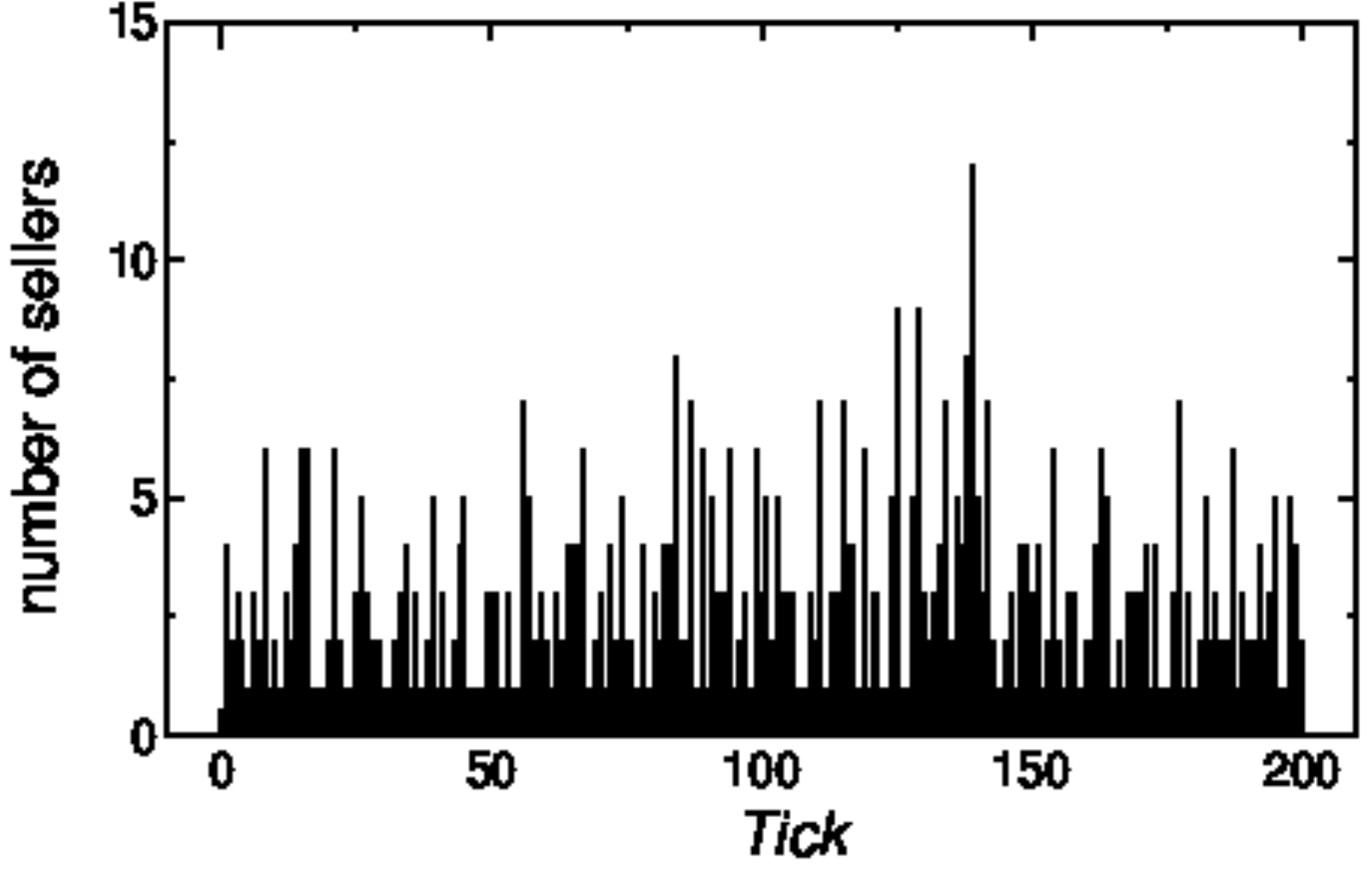} \\
(b) 
\caption{The number of sellers of Fig.~\ref{fig:behaviours_of_models92}.
         The data is generated by dealer model~92 of of Eq.~(\ref{eq:dealer_model92}).
         Plot~(b) is an enlargement of~(a).
         All other conditions are the same in Fig.~\ref{fig:behaviours_of_models92}.}
    \label{fig:sellers_number92}
%	Fig.~\ref{fig:sellers_number92}
\end{figure}

It is unlikely and unnatural for sellers to know the precise number 
of sellers in any deal in advance (no dealer actually knows it).
However, it is possible to know the number of sellers in the past deals.
Let us examine the behaviours of the number of sellers of the data
shown in Fig.~\ref{fig:behaviours_of_models92}.
Figure~\ref{fig:sellers_number92} shows that the number of sellers
varies wildly at each deal, where the minimum is 1 and the maximum is 14.
It does not seem to be easy to predict the next number of
sellers using the past data.\footnote{We apply 
    the small-shuffle surrogate~(SSS) method
    to the data~\cite{Nakamura-Small:SSS_PRE}.
    The result indicates that the data are independently distributed random 
    variables or time-varying random variables.
    Hence, we conclude that we cannot predict the number of sellers using the past data.}
In such cases each seller would have
no other choice than to obey his/her own guess for the next price change.
As a result, the expectation for buying and selling would not always cancel out.
To formulate this situation simply,
we use the average seller number~$ \mu_n(t) $
instead of the precise seller number~$ n $ in Eq.~(\ref{eq:Delta}), 
where $ \mu_n(t) $ is calculated using all of the seller number
prior to time~$ t $ (that is, from $ 1 $ to $ t-1 $).
We note that this idea is not so irrelevant.
When the prediction of the future value does not seem to be easy,
one of the common approaches is to use the mean of the past values:
the average number of sellers who participated deals in the past.
Based on these considerations,\footnote{The essence of this modification is 
    to make an imbalance between the sellers' price rises and 
    the buyer's price decline.
    Although the buyer's price decline~$ -\delta $ is a constant value
    in Eq.~(\ref{eq:Delta_with_mu}), 
    the buyer could change the value depending on situations
    and it is indeed possible to use a time dependent or 
    time varying value (that is, $ -\delta(t) $).
    However, we use $ -\delta $ for simplification.}
Eq.~(\ref{eq:Delta}) is modified as follows: 
%-----------------------------------
\begin{eqnarray}
    \Delta_i = 
    \begin{cases}
    -\delta & \text{(for the buyer)}, \\
    \delta/\mu_n(t) & \text{(for the sellers)}, \\
    0 & \text{(for the nonparticipants of the deal)},
    \end{cases}
    \label{eq:Delta_with_mu}
%	Eq.~(\ref{eq:Delta_with_mu})
\end{eqnarray}
%-----------------------------------
where $ 0 < \delta < L $.

%-----------------------------------
%-----------------------------------

\begin{figure}[!t]
\centering
\includegraphics[width=6.5cm]{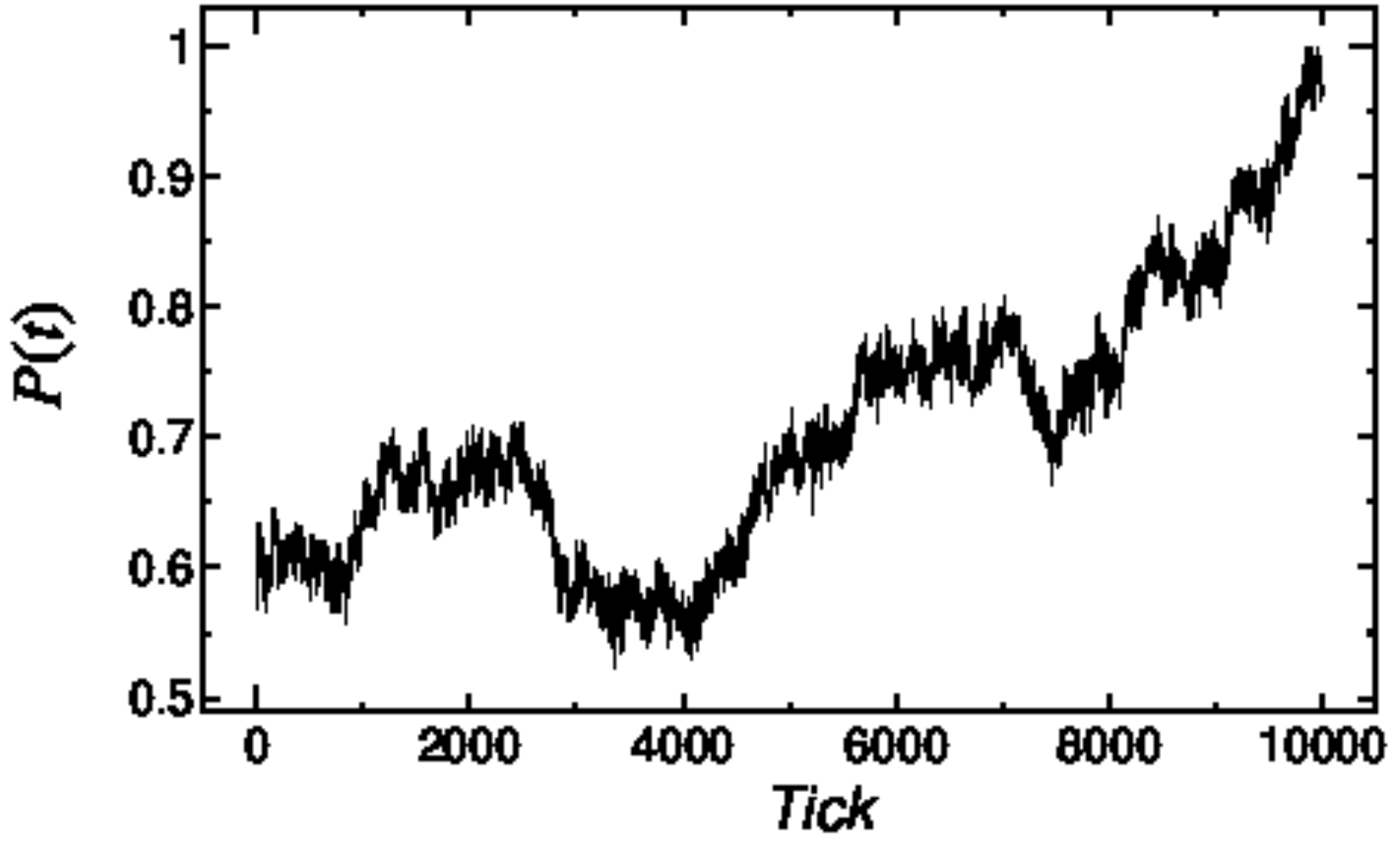} \\
(a) \\
\includegraphics[width=6.5cm]{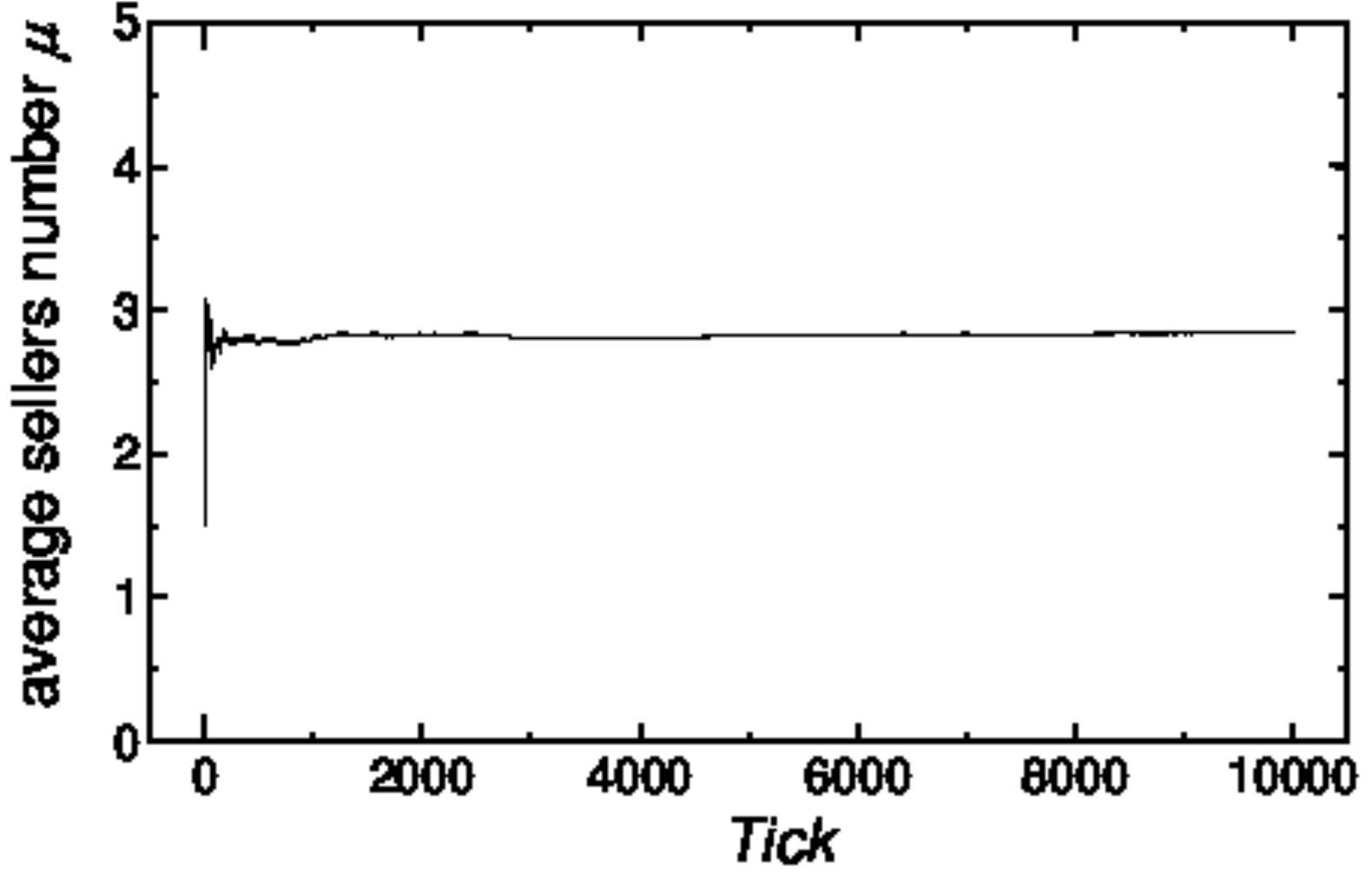} \\
(b) \\
\includegraphics[width=6.5cm]{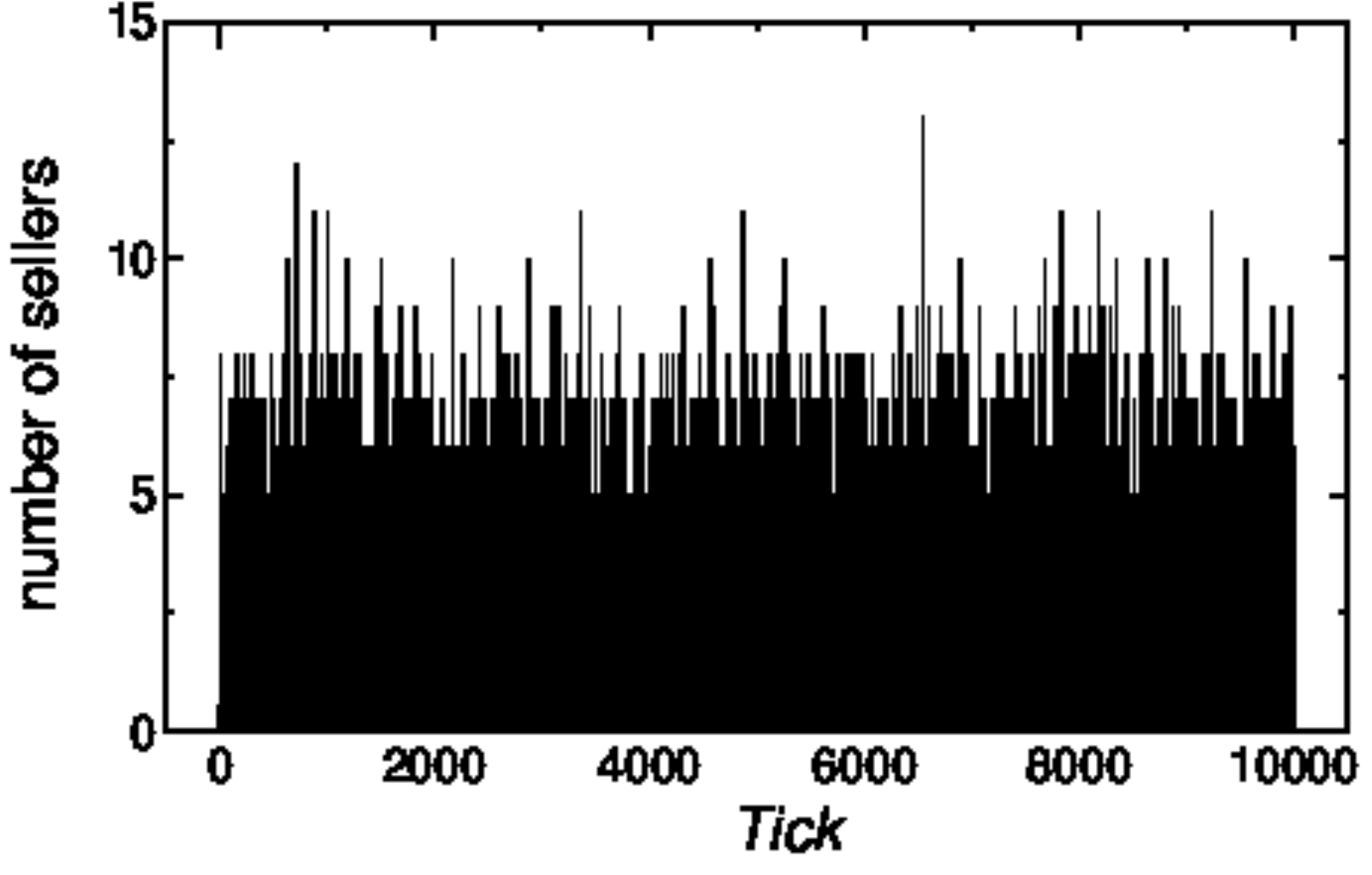} \\
(c)
\caption{Typical behaviours of the dealer models~92 using the average number 
         of sellers in the past.
         (a)~simulated price data~$ P(t) $, 
         (b)~the average number of sellers,
         and (c)~the number of sellers at each deal.
         All other parameter settings are the same in Fig.~\ref{fig:behaviours_of_models92}.}
    \label{fig:dealer_model92_with_trends}
%	Fig.~\ref{fig:dealer_model92_with_trends}
\end{figure}

Figure~\ref{fig:dealer_model92_with_trends} shows the behaviours of 
the dealer models~92 with the modification by Eq.~(\ref{eq:Delta_with_mu}),
where the average sellers number is used instead of the precise sellers number.
Figure~\ref{fig:dealer_model92_with_trends}(a) shows that 
the simulated price data~$ P(t) $ have short-term variabilities and 
exhibit the appearance of long-term rising and falling behaviours of trends.
The difference in the behaviour from the one for
the original dealer model 92 shown in Fig.~\ref{fig:behaviours_of_models92}
due to the simple modification 
from $ n $ to $ \mu_n(t) $ in the definition of $ \Delta_i $ is striking.
The overall behaviour seems to be similar to the real data
shown in Fig.~\ref{fig:financial_data_with_trends}.
The original dealer model~92 with Eq.~(\ref{eq:Delta})
generates only behaviours without trends~\cite{Takayasu-etal:dealer_model92},
as shown in Fig.~\ref{fig:behaviours_of_models92}.
Although we do not input previously prepared trends and 
add any driving force term to generate trends forcibly to the model,
a variety of trends emerge under the condition
that complete information about deals cannot be obtained in advance.
This result implies that the dealer model~92 
intrinsically contains the ability to behaviour trends
without adding any external terms.
Figure~\ref{fig:dealer_model92_with_trends}(b) shows that 
the average number of sellers converges very quickly.
In contrast, from Fig.~\ref{fig:dealer_model92_with_trends}(c)
that shows the number of sellers 
at each deal,\footnote{We apply the SSS method to the number of sellers 
    (not average number)~\cite{Nakamura-Small:SSS_PRE}.
    As it is for the previous application of the SSS method to the average number,
    the result indicates again that the number of sellers are
    also independently distributed random variables 
    or time-varying random variables.}
the number fluctuates wildly at each deal between 1 and 13.
The behaviour is similar to Fig.~\ref{fig:sellers_number92}(a).
We emphasize that the emergence of the long-term monotonic increase
and decrease of price change (that is, ``trends'')
from the dealer model~92 is rarely reported.
In Fig.~\ref{fig:dealer_model92_with_trends} we calculate $ \mu_n(t) $ 
using all of the seller number data prior to the deal at $t$,
although taking the average over all seller number is not essential.
Averaging over the data in a certain interval, that is to say,
the last 100 seller number data from $ t-1 $ to $ t-100 $, 
is equally permissible for generating trends.
%-----------------------------------
%-----------------------------------

As mentioned above, 
the dealer model~92 with the present modification, Eq.~(\ref{eq:Delta_with_mu}),
includes no driving force term to generate previously prepared trends.
% and previously prepared trends are not input to the model.
This result suggests that
trends are not generated dynamically by external driving forces
but emerge spontaneously from the aggregation of microscopic expectations of 
individual dealers.  

%-----------------------------------
%	\setcounter{figure}{9}
\begin{figure}[!t]
\centering
\subfigure[]{\includegraphics[width=6.5cm]{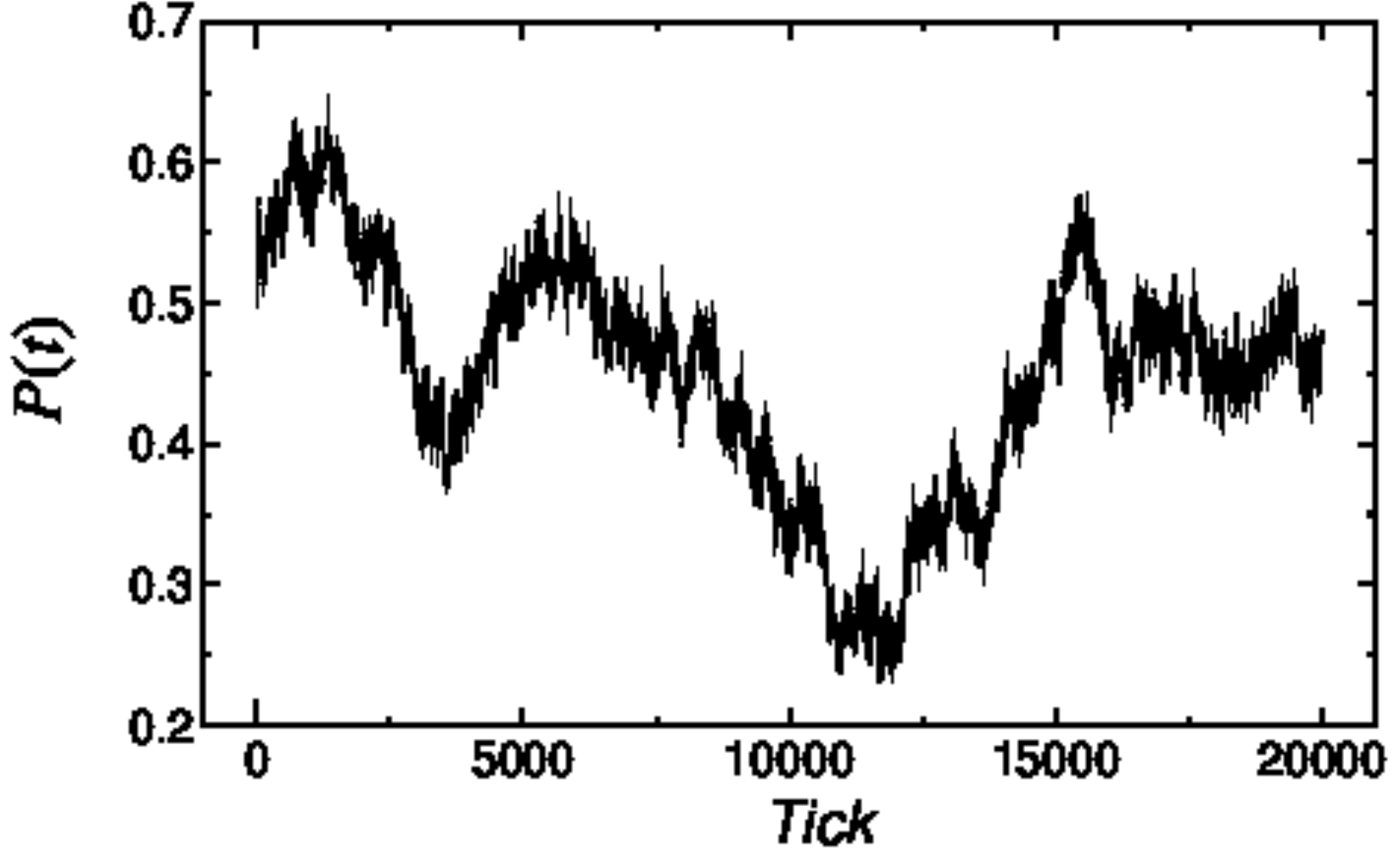}} 
\subfigure[]{\includegraphics[width=6.5cm]{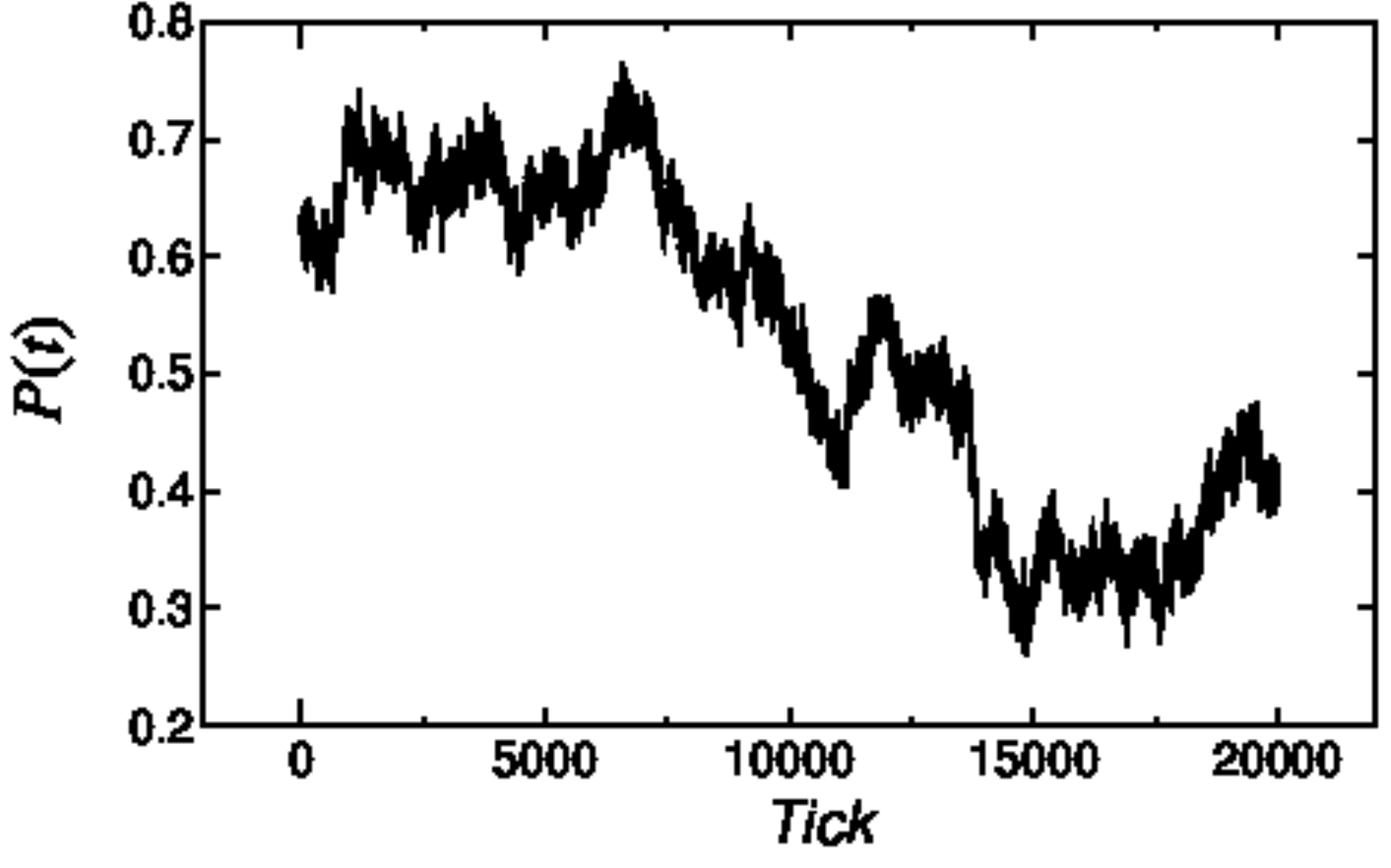}}
\caption{Different behaviours from that shown in 
         Fig.~\ref{fig:dealer_model92_with_trends}(a),
         where different initial conditions of $ B_i $ and 
         different values of $ a_i $ are used.
         All other parameter settings are the same in Fig.~\ref{fig:dealer_model92_with_trends}.}
    \label{fig:different_dealer_model92_with_trends}
%	Fig.~\ref{fig:different_dealer_model92_with_trends}
\end{figure}
%-----------------------------------

Figure~\ref{fig:different_dealer_model92_with_trends} shows 
different behaviours from that shown in Fig.~\ref{fig:dealer_model92_with_trends}(a),
where different initial conditions of $ B_i $ and different values of $ a_i $ 
are used.
We find that there are short-term variabilities and trends
in Figs.~\ref{fig:different_dealer_model92_with_trends}(a) and (b).
We can also recognize a variety of movements such as upward, downward or sideways
in Figs.~\ref{fig:different_dealer_model92_with_trends}(a) and (b).
A trend in a long-term includes finer trends in shorter terms.  
For example, in Fig.~\ref{fig:different_dealer_model92_with_trends}(b)
we recognize downward trends around the ticks
between 7000 and 8500 and between 9000 and 11000, 
and an upward trend between 11000 and 12000.
When we look at these price changes in a longer time scale,
we recognize a downward trend around between 7000 and 15000.
The overall downward trend thus contains these three short-term trends.

We also find another feature.
One of the characteristic changes in the market price evolution 
is its sharp or abrupt change~\cite{Takayasu-etal:dealer_model92}.
We find it around the ticks between 25500 and 26000
on Fig.~\ref{fig:financial_data_with_trends}(a)
and between 24900 and 25500 on Fig.~\ref{fig:financial_data_with_trends}(b).
We also find a similar sharp change on 
Fig.~\ref{fig:different_dealer_model92_with_trends}(b) between 13700 and 13800.
It is remarkable that a little modification of
the original dealer model~92 by Eq.~(\ref{eq:Delta_with_mu}) can produce
such a variety of price changes exhibiting vital features seen
in real financial markets.

%%%%%%%%%%%%%%%%%%%%%%%%%%%%%%%%%%%%%%%%
\subsection{Mingled trends with supposition and no supposition: 
            premeditated trends and unpremeditated trends}
\label{subsec:mingled_trends}
%	Section~\ref{subsec:mingled_trends}

In Section~\ref{subsec:trends_with_supposition}, we discussed an idea 
to generate monotonic increase and decrease trends.
In Section~\ref{subsec:trends_with_no_supposition},
we discussed the emergence of trends due to 
the imbalance between the momentum of buying and selling
because of incomplete information about the next deal.
Although we discussed these two cases separately to make the
arguments clear, they are strongly interconnected and 
mingled with each other in real financial markets.
In this Section, therefore, we treat these two cases on the same footing.

Incorporating both ideas described in Sections~\ref{subsec:trends_with_supposition}
and \ref{subsec:trends_with_no_supposition},
we adopt the following combined and improved modification for $ \Delta_i $:
%-----------------------------------
\begin{eqnarray}
    \Delta_i = 
    \begin{cases}
    -\delta \times (1.0 + \varepsilon_{b}) & \text{(for the buyer)}, \\
    \delta / \mu_n(t) & \text{(for the sellers)}, \\
    0 & \text{(for nonparticipants of the deal)},
    \end{cases}
    \label{eq:Delta_with_epsilon&mu}
%	Eq.~(\ref{eq:Delta_with_epsilon&mu})
\end{eqnarray}
%-----------------------------------
where the notations of $ \delta $ and $ n $ are the same as described 
in Eq.~(\ref{eq:Delta}) and the notations of $ \varepsilon_{b} $ 
and $ \mu_n(t)$ as the same as described in
Eqs.~(\ref{eq:Delta_with_epsilon}) and (\ref{eq:Delta_with_mu}).

\begin{figure}
%	\begin{figure}[!h]
%	\begin{figure}[!t]
\centering
{\includegraphics[width=6.5cm]{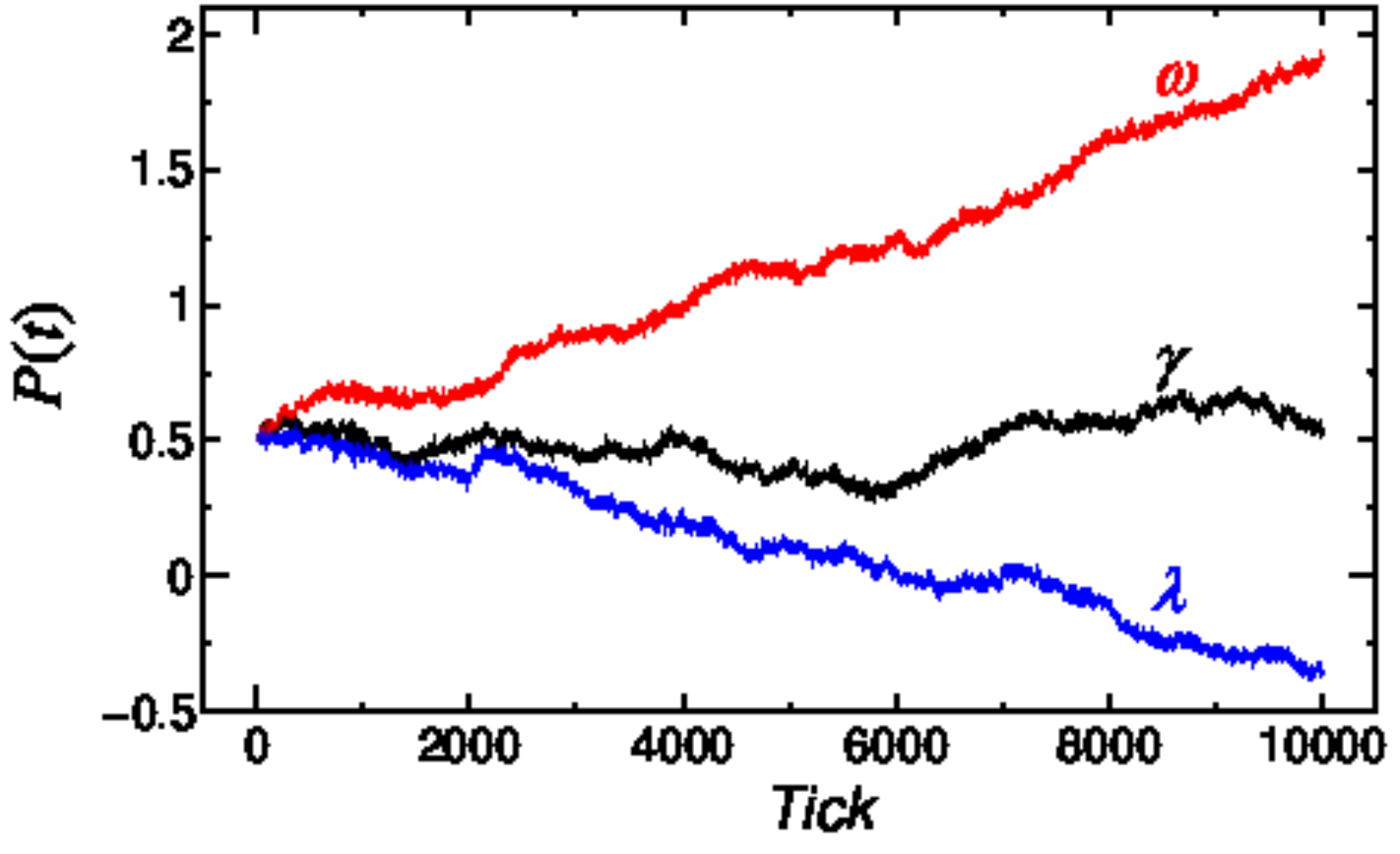}} \\
(a) \\
{\includegraphics[width=6.5cm]{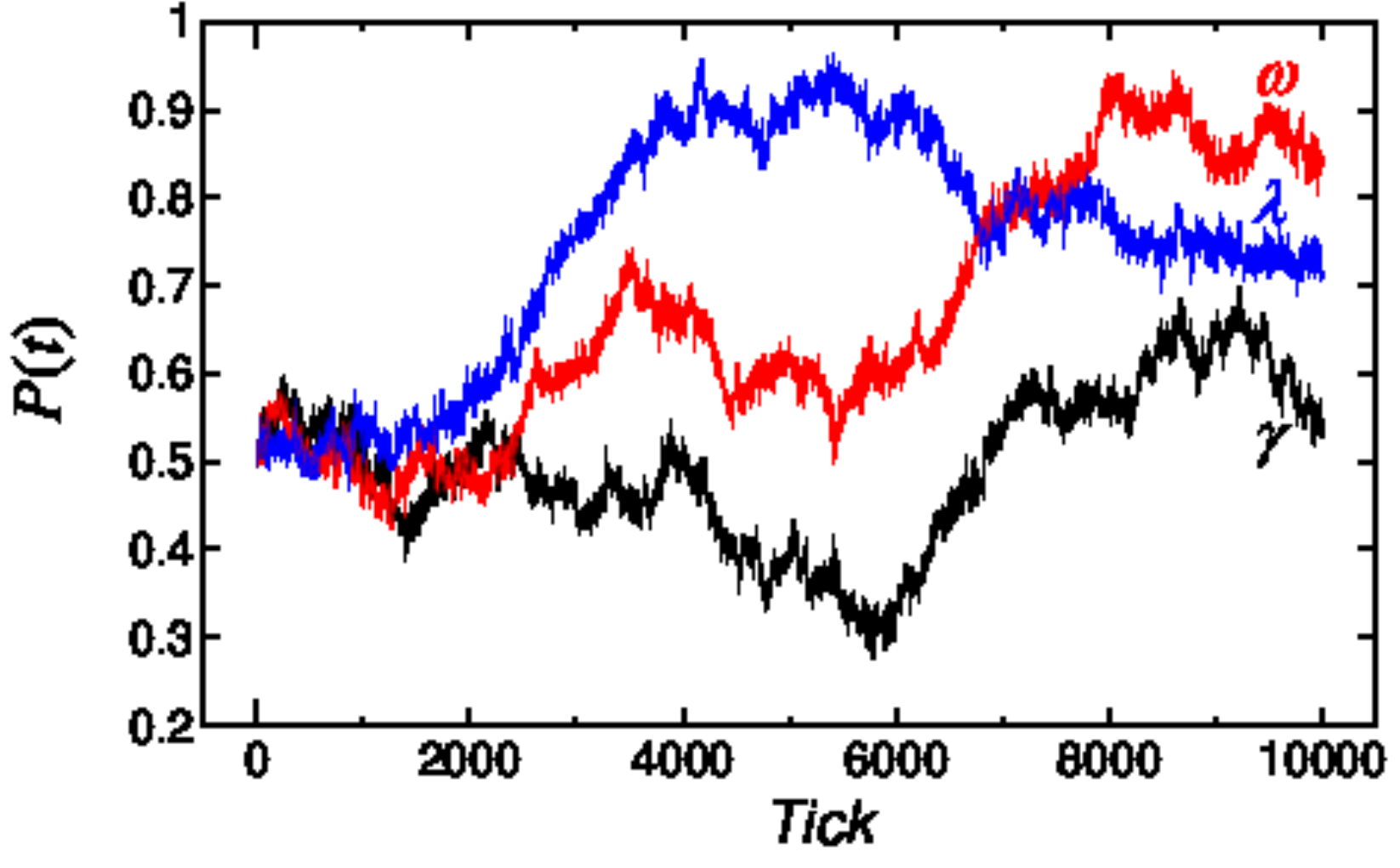}} \\
(b) \\
{\includegraphics[width=6.5cm]{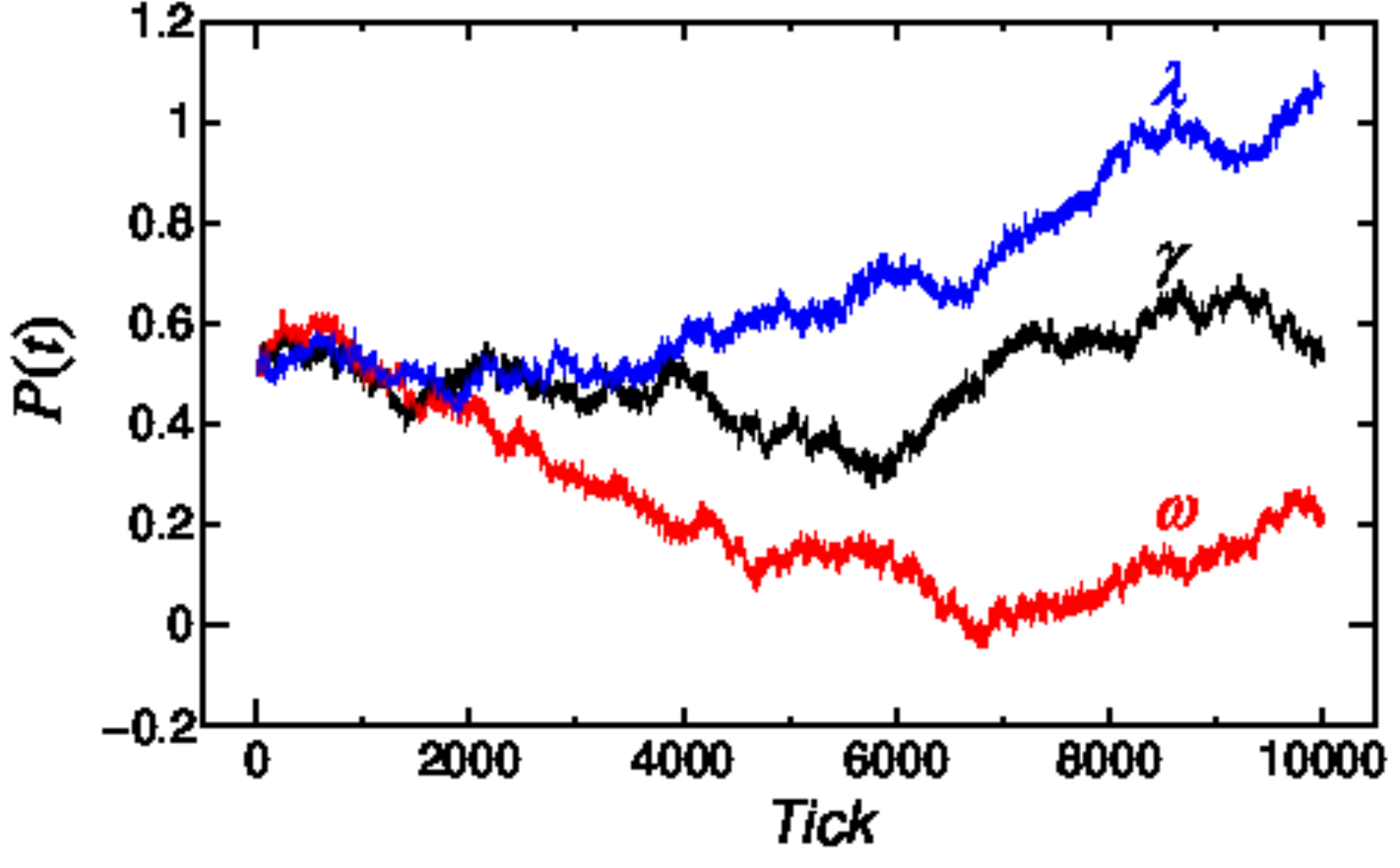}} \\
(c) 
\caption{(Colour online)~Mingled trends generated by
         the improved dealer model~92
         defined by Eq.~(\ref{eq:Delta_with_epsilon&mu}):
         (a)~$ \varepsilon_b = -0.031 $ for plot~$ \omega $
         and $ \varepsilon_b =  0.031 $ for plot~$ \lambda $,
         (b)~$ \varepsilon_b = -0.0021 $ for plot~$ \omega $
         and $ \varepsilon_b =  0.0021 $ for plot~$ \lambda $,
         (c)~$ \varepsilon_b = -0.002 $ for plot~$ \omega $
         and $ \varepsilon_b =  0.002 $ for plot~$ \lambda $.
         All plots labelled by $ \gamma $ in (a), (b), and (c)
         are the data generated by $ \varepsilon_b = 0 $.
         Note that
         plot of~$ \omega $ originally corresponds to a monotonic increase trend,
         that of~$ \lambda $ to a monotonic decrease trend,
         and that of~$ \gamma $ to unpremeditated trends.
         All other conditions are the same in Fig.~\ref{fig:behaviours_of_models92}.}
    \label{fig:dealer_model92_with_trends&monotonic_change}
%	Fig.~\ref{fig:dealer_model92_with_trends&monotonic_change}
\end{figure}

We generate price changes using the dealer model~92 with the modification
Eq.~(\ref{eq:Delta_with_epsilon&mu}) 
for different values of $ \varepsilon_b $.
The results show complicated and a wide variety of behaviours.
Figure~\ref{fig:dealer_model92_with_trends&monotonic_change} shows 
one of the explicit results.\footnote{We examined 
    the behaviours using different values of initial conditions of $ B_i $,
    that of $ a_i $ and that of $ \varepsilon_b $.
    We found that remarkable different behaviours were found 
    at $ |\varepsilon_b| = 0.015 $.
    When $ |\varepsilon_b | >= 0.015 $ we found that monotonic increase or 
    decrease trends tend to appear.
    When $ |\varepsilon_b |< 0.015 $ we found complicated behaviours 
    like Figs.~\ref{fig:dealer_model92_with_trends&monotonic_change}(b) and (c).}
We also show the plots for the data corresponding to
unpremeditated trends defined by $ \varepsilon_b = 0 $ for comparison.
By replacing the number of sellers~$ n $ with
the average value $ \mu_n(t) $,
the profiles of the trends becomes significantly
susceptible to the actual values of $ \varepsilon_b $.
Without the replacement of $ n $ with $ \mu_n(t) $,
a negative $ \varepsilon_b $ generates a monotonically increasing trend,
and a positive $ \varepsilon_b $ generates a monotonically decreasing trend
as shown in Fig.~\ref{fig:monotonic_change92}.
Figure~\ref{fig:dealer_model92_with_trends&monotonic_change}(a) shows 
similar results for $ \varepsilon_b = \pm 0.031 $.
However, the sign of $ \varepsilon_b $ does not have a simple relationship
with increasing or decreasing nature of trends in this case.
For Fig.~\ref{fig:dealer_model92_with_trends&monotonic_change}(b),
we use a much smaller absolute value for $ |\varepsilon_b| = 0.0021 $
than that used in Fig.~\ref{fig:dealer_model92_with_trends&monotonic_change}(a),
$ |\varepsilon_b| = 0.031 $.
In this plot, a trend produced by the positive value, $ \varepsilon_b = 0.0021 $,
shows much more prominent increase than that produced by the negative value,
$ \varepsilon_b = -0.0021 $.
Only by taking a slightly smaller absolute value,
$ |\varepsilon_b| = 0.002 $, we see that a negative value,
$ \varepsilon_b = -0.002 $, generates a decreasing trend
and a positive value, $ \varepsilon_b = 0.002 $, generates an increasing trend
as shown in Fig.~\ref{fig:dealer_model92_with_trends&monotonic_change}(c).
Note that these values, $ |\varepsilon_b| = 0.002 $, are the same as 
those used in Fig.~\ref{fig:monotonic_change92}.
%-----------------------------------
%-----------------------------------
%%%%%%%%%%%%%%%%%%%%%%%%%%%%%%%%%%%%%%%%
\section{Discussions and Summary}
\label{sec:discussion&summary}
%	Section~\ref{sec:discussion&summary}

When we use both ideas to generate trends with supposition
and no supposition at the same time in Section \ref{subsec:mingled_trends},
we observed two distinctive aspects.
One is that dealers' minuscule tendencies in next price setting can make
a large difference in the overall behaviours of price change.
The other is that there are cases where the behaviours of price change 
are different from, even opposite to, individual dealers' expectations.
% of price change.
Both of the aspects indicate that the direction of trends cannot tell us 
the underlying expectations of the dealers.
In other words, the behaviours of the trend do not always accord with 
dealers' long-term expectations.
This leads to a very intriguing conclusion. 
From the behaviours or movements of price change,
each dealer infers his/her specious justification
of the cause of the change and offer future prospects
to others in the next deal.
Once the price in a market moves against their expectations, however,
dealers doubt their justifications
and may take a counter reaction to their original expectation.
Such counter reactions of the dealers may cause excess volatility.

We understand that 
there might be other origins to generate trends
than the ones described in this paper.
The reason we believe that the origins discussed in the present paper
are essential is their universality.
As no dealer has access to complete information about the next deal in actual financial markets,
dealers have no choice other than to have their own expectations.
As these expectations are brought by incomplete information,
the expectations would be rough and flimsy.
Such flimsy expectations of individual dealers
break a subtle balance of acquisitive natures between buyers and sellers
and generate macroscopic trends,
even if there are no external driving factors.
Since the situations in which
we cannot obtain complete information about the next status are very common,
we believe that the knowledge
gained from the ideas pursued in this paper contains an essential ingredient
and can be applicable to many other situations
For example, we often find the price changes in actual markets
that exhibit very similar behaviours, although the brands are different,
such as Australian Dollar/USD exchange rate and 
New Zealand Dollar/USD exchange rate.

Our results indicate that some trends might be unpremeditated.
Although the trends emerge spontaneously,
why their behaviours are so similar?
What kind of interactions are there between the brands?
Those will be very interesting future questions.

Summary of the results is the following.
In this paper,
we have investigated possible origins of trends in financial market price changes
using a deterministic threshold model called the dealer model~92.
% from the viewpoint of dealers' expectation for the price.
We consider financial markets as a game field of commercial activity 
composed of human choices and their interactions.
First, we have observed that monotonically increasing and decreasing trends can be generated
by dealers' minuscule tendencies for price setting according
to the expected forthcoming price changes.
Next, we have observed that
irregular fluctuations with short-term variabilities and trends 
emerge spontaneously under the assumption that
we cannot obtain complete information about the next deal in advance.
In both cases, trends are generated without introducing any external driving forces
to the original dealer modes~92.
Finally, we have confirmed that the numerical data for price change generated
by the dealer model~92 incorporating both of these two cases contains several vital
features observed in real financial markets.
These results indicates that emergence of trends
might be inevitable in any realistic situations, 
even if the influence on fundamentals and technical analyses 
in financial markets is almost absent.

%%%%%%%%%%%%%%%%%%%%%%%%%%%%%%%%%%%%%%%%
\section{Acknowledgements}

We would like to thank the anonymous referees for their very valuable remarks.
One of the authors, TT, would like to acknowledge
the support of Grant-in-Aid for Scientific Research (C)
(No.\ 24540419) from the Japan Society for the Promotion of Science.
%%%%%%%%%%%%%%%%%%%%%%%%%%%%%%%%%%%%%%%%%

%%%%%%%%%%%%%%%%%%%%%%%%%%%%%%%%%%%%%%%%%
% References
% \bibliography{bibliography} % Produces the bibliography via BibTeX.

%-----------------------------------------------------------------
\end{document}